\documentclass[11pt,letterpaper]{article}

\usepackage{thmtools}
\usepackage{thm-restate}

\usepackage{mathrsfs}           
\usepackage{vmargin,fancyhdr}   
\usepackage{algorithm}
\usepackage{algorithmic}

\usepackage{enumerate}

\usepackage{amsmath,amssymb}    
\usepackage{verbatim}           
\usepackage{xspace}             
\usepackage{graphicx,float}     
\usepackage{ifthen,calc}        
\usepackage{textcomp}           
\usepackage{fancybox}           
\usepackage{hhline}             
\usepackage{float}              

\usepackage[rflt]{floatflt}     
\usepackage[small,compact]{titlesec}
\usepackage{setspace}
\usepackage{subfigure}

\usepackage{color}
\definecolor{ForestGreen}{rgb}{0.1333,0.5451,0.1333}
\usepackage[letterpaper,
            colorlinks,linkcolor=ForestGreen,citecolor=ForestGreen,
            backref,
            bookmarks,bookmarksopen,bookmarksnumbered]
           {hyperref}
\usepackage{thumbpdf}

\setpapersize{USletter}
\setmarginsrb{.75in}{.5in}        
             {.75in}{.5in}        
             {.25in}{.25in}     
             {.25in}{.5in}      
\newcommand{\showccc}[0]{0}
\newcommand{\ccc}[2][nothing]{
  \ifthenelse{\showccc=0}{}{
    \ensuremath{^{\Lsh\Rsh}}\marginpar{\raggedright\tiny\textsf{%
        \ifthenelse{\equal{#1}{nothing}}{}{\textbf{#1}\\}#2}}}}

\newcounter{hours}\newcounter{minutes}
\newcommand{\hhmm}{%
  \setcounter{hours}{\time/60}%
  \setcounter{minutes}{\time-\value{hours}*60}%
  \ifthenelse{\value{hours}<10}{0}{}\thehours:%
  \ifthenelse{\value{minutes}<10}{0}{}\theminutes}
\ifthenelse{\showccc=0}{\rhead{}}{\rhead{\today \ [\hhmm]}}
\newtheorem{theorem}{Theorem}[section]
\newtheorem{claim}[theorem]{Claim}

\newtheorem{corollary}[theorem]{Corollary}
\newtheorem{definition}[theorem]{Definition}

\newtheorem{lemma}[theorem]{Lemma}
\newtheorem{fact}[theorem]{Fact}

\newcommand{\Proof}[0]{\smallskip\noindent\textit{\textbf{Proof}}\quad}
\newcommand{\Proofof}[1]{\smallskip\noindent\textit{\textbf{Proof of #1:}}\quad}

\newcommand{\QED}[0]{\hfill\ensuremath{\blacksquare}\medspace}

\begin{document}

\title{Preconditioning in Expectation}

\author{
  Michael B. Cohen\\
  M.I.T.\thanks{Part of this work was done while at CMU}\\
  \texttt{micohen@mit.edu}
  \and
  Rasmus Kyng\\
  Yale University\thanks{This work was partially supported by AFOSR Award FA9550-12-1-0175.} \\
  \texttt{rasmus.kyng@yale.edu}
  \and
  Jakub W. Pachocki\\
  Carnegie Mellon University\\
  \texttt{pachocki@cs.cmu.edu}
  \and
  Richard Peng\\
  M.I.T.\thanks{Part of this work was done while at CMU and was supported by a Microsoft Research PhD Fellowship}\\
  \texttt{rpeng@mit.edu}
  \and
  Anup Rao\\
  Yale University \thanks{This work was partially supported by NSF grant CCF-1111257.} \\
  \texttt{anup.rao@yale.edu}
}

\def\bvec#1{{\mbox{\boldmath $#1$}}}

\newcommand\bb{\boldsymbol{\mathit{b}}}
\newcommand\bbhat{\widehat{\boldsymbol{\mathit{b}}}}
\newcommand\dd{\boldsymbol{\mathit{d}}}
\newcommand\ff{\boldsymbol{\mathit{f}}}
\newcommand\rr{\boldsymbol{\mathit{r}}}
\newcommand\xx{\boldsymbol{\mathit{x}}}
\newcommand\yy{\boldsymbol{\mathit{y}}}
\newcommand\xxbar{\overline{\boldsymbol{\mathit{x}}}}
\newcommand\xxhat{\widehat{\boldsymbol{\mathit{x}}}}
\newcommand\yybar{\overline{\boldsymbol{\mathit{y}}}}

\newcommand\ww{\boldsymbol{\mathit{w}}}

\newcommand{\erExact}{\ensuremath{\overline{\boldsymbol{\tau}}}}
\newcommand{\er}{\ensuremath{\boldsymbol{\tau}}}

\renewcommand\AA{{\mathit{A}}}
\newcommand\BB{{\mathit{B}}}
\newcommand\HH{{\mathit{H}}}
\newcommand\II{{\mathit{I}}}
\newcommand\LL{{\mathit{L}}}
\newcommand\MM{{\mathit{M}}}
\newcommand\PP{{\mathit{P}}}
\newcommand\RR{{\mathit{R}}}
\newcommand\UU{{\mathit{U}}}
\newcommand\WW{{\mathit{W}}}
\newcommand\XX{{\mathit{X}}}
\newcommand\YY{{\mathit{Y}}}
\newcommand\ZZ{{\mathit{Z}}}

\newcommand\PPi{{\Pi}}

\def\norm#1{\left\| #1 \right\|}

\newcommand\eFlow[1]{\mathcal{E}_{\RR}(#1)}

\newcommand{\defeq}{\buildrel \text{d{}ef}\over =}
\newcommand{\Oh}{\ensuremath{\mathcal{O}}}
\newcommand{\oh}{\ensuremath{\mathcal{O}}}
\newcommand{\hsum}[2]{\operatorname{HrmSum}{\ensuremath{\left(#1, #2\right)}}}

\def\prob#1#2{\mbox{Pr}_{#1}\left[ #2 \right]}
\newcommand{\expct}[2]{\ensuremath{\mathbb{E}}_{#1}\left[#2\right]}
\newcommand{\nbr}[1]{\left\|#1\right\|}
\newcommand{\trace}[1]{\operatorname{Tr}\left( #1 \right)}
\newcommand{\func}[1]{f_{#1}}
\newcommand{\cexpct}[3]{\ensuremath{\mathbb{E}}_{#1}\left[#2\vert#3\right]}

\newcommand{\poly}{\ensuremath{\textbf{poly}}}

\newcommand{\mati}{\ensuremath{I}}
\newcommand{\laplacian}{\ensuremath{L}}
\newcommand{\mata}{\ensuremath{A}}
\newcommand{\matb}{\ensuremath{B}}
\newcommand{\matx}{\ensuremath{X}}
\newcommand{\maty}{\ensuremath{Y}}
\newcommand{\matz}{\ensuremath{Z}}
\newcommand{\matnz}{\ensuremath{W}}
\newcommand{\mathh}{\ensuremath{H}}
\newcommand{\matnx}{\ensuremath{S}}
\newcommand{\matny}{\ensuremath{R}}
\newcommand{\matni}{}

\newcommand{\vecindicator}{\ensuremath{\boldsymbol{\chi}}}
\newcommand{\vecb}{\ensuremath{\textbf{b}}}
\newcommand{\err}{\ensuremath{\textbf{err}}}
\newcommand{\vecr}{\ensuremath{\textbf{r}}}
\newcommand{\vecnv}{\ensuremath{\textbf{u}}}
\newcommand{\vecnnv}{\ensuremath{\textbf{z}}}
\newcommand{\vecv}{\ensuremath{\textbf{v}}}
\newcommand{\vecx}{\ensuremath{\textbf{x}}}
\newcommand{\vecy}{\ensuremath{\textbf{y}}}

\maketitle

\begin{abstract}
We show that preconditioners constructed by random sampling can perform
well without meeting the standard requirements of iterative methods.
When applied to graph Laplacians, this leads to ultra-sparsifiers that in expectation
behave as the nearly-optimal ones given by [Kolla-Makarychev-Saberi-Teng STOC`10].
Combining this with the recursive preconditioning framework by [Spielman-Teng STOC`04] and improved embedding algorithms,
this leads to algorithms that solve symmetric diagonally dominant linear systems
and electrical flow problems in expected time close to $m\log^{1/2}n$ .
\end{abstract}

\thispagestyle{empty}
\newpage

\section{Introduction}
\label{sec:intro}

Randomized constructions of algebraically similar objects are widely
used in the design of efficient algorithms.
Sampling allows one to reduce the size of a problem while preserving
its structure, and then solve the problem on a smaller instance.
It is a core component in randomized matrix algorithms~\cite{Mahoney11},
stochastic gradient descent~\cite{Bottou04},
and graph algorithms.

Smaller equivalents of graphs are known as \emph{sparsifiers}, and the study of
sampling methods for generating them led to the cut sparsifiers by
Benczur and Karger~\cite{BenczurK96}, and spectral sparsifiers
by Spielman and Teng~\cite{SpielmanTengSparsifier}.
Spectral sparsifiers are key routines in the first nearly-linear time solver
by Spielman and Teng~\cite{SpielmanT04}, as well as in the subsequent
improvements by Koutis et al.~\cite{KoutisMP10,KoutisMP11}.
These solvers, in turn, have many applications which are described
in detail in surveys by Spielman~\cite{Spielman10} and Teng~\cite{Teng10}.

At the core of the Spielman and Teng solver is a recursive preconditioning 
framework which transfers solutions between a sequence of sparsifiers
known as a solver chain.
Improvements to this framework led to algorithms that run in
 about $m\log{n}$ time under exact arithmetic~\cite{KoutisMP11}.
The existence of an algorithm that solves a given system in
about $m\log^{1/2}n$ time after preprocessing can be derived from
the nearly-optimal \emph{ultra-sparsifiers} by Kolla et al.~\cite{KollaMST10}.
These ultra-sparsifiers build upon the nearly-optimal
spectral sparsifiers by Batson et al.~\cite{BatsonSS09},
and gain a factor of $\log^{1/2}{n}$ over randomized constructions.
However, the current fastest algorithm for constructing these objects
by Zouzias~\cite{Zouzias12} takes cubic time.
As a result, finding nearly-linear time algorithms for constructing
nearly-optimal sparsifiers and ultra-sparsifiers were posed as
an important open question in the article by Batson et al.~\cite{BatsonSST13}.

Recently, a new approach to solving SDD linear systems was proposed by
Kelner et al.~\cite{KelnerOSZ13},
and extended by Lee and Sidford~\cite{LeeS13}.
Instead of constructing spectral sparsifiers, they show that fixing single cycles chosen
from an appropriate distribution leads to sufficient decreases in errors \emph{in expectation}.
In this paper, we extend this approach to more general subgraphs, and show that this
 achieves the same improvement per iteration as the optimal ultra-sparsifiers, \emph {in expectation}.
Our results can therefore be viewed as an algorithmic answer to the open question by
Batson et al.~\cite{BatsonSST13} on efficiently generating nearly-optimal sparsifiers.

Similar to the spectral sparsifiers by Batson et al.~\cite{BatsonSS09},
our results are applicable to general matrices.
Instead of aiming to show that the sampled matrix is a good spectral approximation,
our analysis is geared towards the intended application of the sample:
use as a preconditioner for iterative methods for solving linear systems.
We discuss these iterative methods and the statistical bounds needed for
their convergence in Section~\ref{sec:overview}.
This randomized iterative method resembles to the randomized block
Kaczmarz method by Needell and Tropp~\cite{NeedellT13}.
However, our convergence guarantees are more akin to those of
standard iterative methods such as the ones presented in~\cite{Axelsson94:book}.

For linear systems in Laplacians of graphs, our randomized iterative
methods can be incorporated into existing solver frameworks.
In Section~\ref{sec:sdd}, we use the recursive preconditioning framework by
Koutis et al.~\cite{KoutisMP11} to obtain the following result:
\begin{theorem}
\label{thm:mainSDD}
Given a graph $G$ with $m$ edges, a vector $\vecb = \laplacian_G \vecx$, and any error $\epsilon > 0$, we can find w.h.p. a vector $\vecx$ such that
\begin{align*}
	\nbr{\bar{\vecx} - \vecx}_{\laplacian_G} \leq \epsilon \nbr{\bar{\vecx}}_{\laplacian_G},
\end{align*}
in expected $\Oh(m\log^{1/2}n \log\log^{3+\delta}n \log(\frac{1}{\epsilon}))$ time for any constant $\delta > 0$.
\end{theorem}
In Appendix~\ref{sec:electrical}, we show that this solver can also be used to
generate electrical flows with approximate minimum energy in similar time.
This problem is dual to solving linear systems, and is the core problem
addressed by previous solvers that reduce distance in expectation~\cite{KelnerOSZ13,LeeS13}.

Our presentation of the solver in Section~\ref{sec:sdd} aims for simplicity,
and we do not optimize for the exponent on $\log\log{n}$.
This allows us to reduce to situations where errors of $\poly{\log{n}}$
can be tolerated.
Here we can use existing algorithms that are guaranteed to
return good answers with high probability.
We believe that this algorithmic dependency is removable, and that the exponent on $\log\log{n}$
can be reduced, or even entirely removed by a more refined analysis.

We also assume that all arithmetic operations are exact in this paper.
The iterative methods used in our algorithm, namely the preconditioned
Chebyshev iteration in Appendix~\ref{sec:cheby}, are stated with robust
bounds that can absorb large absolute error.
Therefore, only the Gaussian elimination stages need to be checked
to show the numerical stability of our algorithm in the setting of fixed-point arithmetic.
Such an analysis of the recursive preconditioning framework can be found in
Section 2.6 of~\cite{Peng:thesis}, and should be readily applicable to our algorithm as well.

\section{Overview}
\label{sec:overview}

Our starting point is the simplest iterative method,
known as Richardson iteration.
In the setting that we use it in, it can also be viewed as iterative refinement.
If our goal is to solve a linear system $\YY \xx = \bb$,
and we have a matrix $\ZZ$ that's similar to $\YY$,
this method generates a new $\xx'$ using the step
\begin{equation}
\label{eq:step}
\xx' = \xx - \alpha \ZZ^{-1} \left( \YY \xx - \bb \right).
\end{equation}

Here $\alpha$ is a parameter that we can choose based on the
approximation factor between $\ZZ$ and $\YY$.
When $\ZZ$ is an exact approximation, i.e. $\ZZ = \YY$, we can set
$\alpha = 1$ and obtain
\[
\xx' = \xx - \YY^{-1} \left( \YY \xx - \bb \right)
= \xx - \xx + \YY^{-1}\bb = \YY^{-1}\bb.
\]

Of course, in this situation we are simply solving $\ZZ \xx = \bb$ directly.
In general, iterative methods are used when $\ZZ$ is an approximation of $\YY$.
The quality of this approximation can be measured using
relative condition numbers, which are defined using spectral orderings.
While our main algorithm relies on a weaker notion of approximation,
this view nonetheless plays a crucial role in its
intermediate steps, as well as its analysis.
Given two matrices $\AA$ and $\BB$, we say $\AA \preceq \BB$
if $\BB - \AA$ is positive semidefinite. Using this ordering,
matrix approximations can then be defined by giving
both upper and lower bounds.
The guarantees of Richardson iteration under this notion of
approximation is a fundamental result in iterative methods~\cite{Axelsson94:book}.
\begin{fact}
\label{fact:convergence}
If $\YY \preceq \ZZ \preceq \kappa \YY$ for some parameter $\kappa$,
and $\xxbar$ is the exact solution satisfying $\YY \xxbar = \bb$,
then taking the step in Equation~\ref{eq:step} with $\alpha = \kappa$
gives:
\[
\norm{\xx' - \xxbar}_{\YY} \leq \left( 1 - \frac{1}{\kappa} \right) \norm{\xx - \xxbar}_{\YY},
\]
\end{fact}
Here $\norm{\cdot}_{\YY}$ is the matrix norm of $\YY$,
$\norm{\cdot}_{\YY} = \sqrt{\xx^T \YY \xx}$.
It is the standard norm for measuring the convergence of iterative methods.

As Equation~\ref{eq:step} requires us to solve a linear system involving $\ZZ$,
it is desirable for $\ZZ$ to be smaller than $\YY$.
One way to do this is to write $\YY$ as a sum of matrices,
$\YY = \sum_{i = 1}^{m}\maty_i$, and pick a subset of these.
This in turn can be done via random sampling.
Here a crucial quantity is the statistical leverage score.
For a matrix $\XX$, the leverage score of $\YY_i$ w.r.t. $\XX$ is
\[
\erExact_i \defeq \trace{\XX^{-1} \YY_i}.
\]
For some $\XX$ and $\YY = \YY_1 + \ldots + \YY_m$, we can generate
a preconditioner $\ZZ$ by sampling a number of $Y_i$s with probabilities
proportional to $\erExact_i$.
We can also use upper bounds on the actual leverage scores, $\er_i$.
The pseudocode for a variant of this routine is given in Figure~\ref{fig:sample}.

\begin{figure}[ht]
\vskip 0.2in
\fbox{
\begin{minipage}{6in}
\noindent $\matz = \textsc{Sample} (\{\maty_1, \ldots, \maty_m\}, X, \er, \delta)$,
\label{line:initZ}
where $\maty_i = \vecv_i \vecv_i^T$ are rank one matrices,
$\er_i$ are upper bounds of leverage scores, $\er_i \geq \erExact_i$ for all $i$,
and $\delta < 1$ is an arbitrary parameter.
\begin{enumerate}
	\item Initialize $\matz$ to $\matx$.
	\item Let $s$ be $\sum_{i = 1}^m \er_i$ and $t=\delta^{-1} s$.
	\item Pick an integer $r$ uniformly at random in the interval $[t,2t-1]$.
	\item For $j = 1 \ldots r$
		\begin{enumerate}
			\item Sample entry $i_j$ with probability proportional to $\er_{i_j}$.
			\item $\matz \leftarrow \matz + \frac{\delta}{\er_{i_j}} Y_{i_j}$.
		\end{enumerate}
	\item Return $Z$.
\end{enumerate}
\end{minipage}
}
\caption{Sampling Algorithm}
\label{fig:sample}
\end{figure}

By applying matrix Chernoff bounds such as the ones by Tropp~\cite{Tropp12},
it can be shown that $\frac{1}{2} \YY \preceq \ZZ \preceq 2 \YY$ when
$\delta$ is set to $\frac{1}{O(\log{n})}$.
We will formalize this connection in Appendix~\ref{sec:whp}.
Scaling the resulting $\ZZ$ by a factor of $2$ then gives a preconditioner 
that can be used to make the step given in Equation~\ref{eq:step}.
The preconditioner produced contains $\XX$ plus $O(s \log{n})$ of the matrices $\YY_i$s.
The Kolla et al.~\cite{KollaMST10} result can be viewed as finding
$\ZZ$ consisting of only $O(s)$ of the matrices,
and $\YY \preceq \ZZ \preceq O(1) \YY$, albeit in cubic time.

Our main result is showing that if we generate $\ZZ$ using \textsc{Sample}
with $\delta$ set to a constant, the step given in Equation~\ref{eq:step}
still makes a constant factor progress in expectation, for an appropriate constant $\alpha$.
We do so by bounding the first and second moments of $\ZZ^{-1}$ w.r.t. $\YY$.
These bounds are at the core of our result.
They are summarized in the following Lemma, and proven in Section~\ref{sec:moments}.

\begin{lemma}
\label{lem:moments}
Suppose $\maty_i = \vecv_i \vecv_i^T$ are rank one matrices with sum $\maty$,
$\matx$ is a positive semidefinite matrix satisfying $\matx \preceq \maty$, $\er_1 \ldots \er_m$ are values that satisfy
$\er_i \geq \trace{X^{-1} Y}$,
and $\delta < 1$ is an arbitrary parameter.
Then the matrix $Z = \textsc{Sample}(\maty_1 \ldots \maty_m, \matx,
\er_1 \ldots \er_m, \delta)$ satisfies:
\begin{enumerate}
\item $\expct{r, i_1 \ldots i_r}{\vecx^T \matz^{-1} \vecx} \leq \frac{1}{1 - 2 \delta} \vecx^TY^{-1}\vecx $, and
\label{part:firstLower}
\item $\expct{r, i_1 \ldots i_r}{\vecx^T \matz^{-1} \vecx} \geq \frac{1}{3} \vecx^T Y^{-1} \vecx^T$, and
\label{part:firstUpper}
\item $\expct{r, i_1 \ldots i_r}{\vecx^T \matz^{-1} \maty \matz^{-1} \vecx} \leq \frac{1}{1 - 3 \delta} \vecx^T \maty^{-1} \vecx^T$.
\label{part:secondUpper}
\end{enumerate}
\end{lemma}

Using these bounds, we can show that an iteration similar to Richardson
iteration reduces errors, in expectation, by a constant factor each step.

\begin{lemma}
\label{lem:richardsonExpct}
Suppose $\XX$ and $\YY$ are invertible matrices such that $\XX \preceq \YY$,
$\bb = \YY \xxbar$, and $\xx$ is an arbitrary vector.
If $\matz = \textsc{Sample} (\maty_1 \ldots \maty_m, X, \er_1 \ldots \er_m, \frac{1}{10})$,
and $\xx'$ is generated using
\[
\xx' = \vecx - \frac{1}{10} \ZZ^{-1} \left( \YY \xx - \bb \right).
\]
Then
\begin{align*}
\expct{r, i_1, i_2, \ldots i_r}{\norm{\xxbar - \xx'}_{\YY}^2}
\leq \left( 1 - \frac{1}{40} \right) \nbr{\xxbar - \xx}_{\YY}^2
\end{align*}
\end{lemma}

\Proof
We first rearrange both sides by substituting in $\vecb = \maty \vecx$,
and letting $\vecy = \bar{\vecx} - \vecx$.
The term in the LHS becomes
\begin{align*}
\bar{\vecx}-\left(\vecx - \frac{1}{10} \matz^{-1} \left( \maty \vecx - \vecb \right) \right)
& = \left( \mati - \frac{1}{10} \matz^{-1} \maty \right) \vecy,
\end{align*}
while the RHS becomes
$\left(1 - \frac{1}{40} \right)  \nbr{\vecy}_{\maty}^2$.

Expanding the expression on the LHS and applying linearity of
expectation gives
\begin{align*}
&\expct{r, i_1, i_2, \ldots i_r}{\nbr{ \left( \mati - \frac{1}{10} \matz^{-1} \maty \right) \vecy}_{\maty}^2}\\
& = \expct{r, i_1, i_2, \ldots i_r}{\vecy^T \maty \vecy
- \frac{2}{10} \vecy^T \maty \matz^{-1} \maty
+ \frac{1}{100} \vecy^T \maty \matz^{-1} \maty \matz^{-1} \maty \vecy}\\
& = \vecy^T \maty \vecy
- \frac{2}{10} \expct{r, i_1, i_2, \ldots i_r}{\vecy^T \maty \matz^{-1} \maty \vecy}
+ \frac{1}{100} \expct{r, i_1, i_2, \ldots i_r}{\vecy^T \maty \matz^{-1} \maty \matz^{-1} \maty \vecy}
\end{align*}
Since $\maty \vecy$ is a fixed vector, we can apply Lemma~\ref{lem:moments}
with it as $\vecv$.
The lower bound on first moment in Part~\ref{part:firstLower} 
allows us to upper bound the first term at
\begin{align*}
 \expct{r, i_1, i_2, \ldots i_r}{\vecy^T \maty \matz^{-1} \maty \vecy}
& \geq \frac{1}{3} \vecy^T \maty  \maty^{-1} \maty \vecy\\
& = \frac{1}{3} \vecy^T \maty \vecy.
\end{align*}
The second term can be upper bounded using Part~\ref{part:secondUpper} 
with the same substitution.
\begin{align*}
 \expct{r, i_1, i_2, \ldots i_r}{\vecy^T \maty \matz^{-1} \maty \matz^{-1} \maty \vecy}
& \leq \frac{1}{1 - 3 \delta} \vecy^T \maty  \maty^{-1} \maty \vecy\\
& = \frac{1}{1 - 3 \delta} \vecy^T \maty \vecy\\
& \leq 2 \vecy^T \maty \vecy,
\end{align*}
where the last inequality follows from the choice of $\delta = \frac{1}{10}$.
Combining these then gives the bound on the expected energy:
\begin{align*}
\expct{r, i_1, i_2, \ldots i_r}{\nbr{\bar{\vecx}
- \left( \vecx - \frac{1}{10} \matz^{-1} \left( \maty \vecx - \vecb \right) \right)}_{\maty}^2}
& \leq \nbr{\vecy}_{\maty}^2 - \frac{2}{30} \vecy^T \maty \vecy
+ \frac{2}{100} \vecy^T \maty \vecy\\
& \leq \left( 1 - \frac{1}{40} \right) \vecy^T \maty \vecy
\end{align*}
\QED

When $\XX$ and $\YY$ are lower rank, we have that $\ZZ$
also acts on the same range space since $\XX$ is added to it.
Therefore, the same bound applies to the case where $\XX$ and $\YY$
have the same null-space.
Here it can be checked that the leverage score of $\YY_i$
becomes $\trace{\XX^{\dag} \YY_i}$, and the step is made
based on pseudoinverse of $\ZZ$, $\ZZ^{\dag}$.
Also, note that for any nonnegative random variable $x$ and moment
$0 < p < 1$, we have $\expct{}{x^{p}} \leq {\expct{}{x}}^{p}$.
Incorporating these conditions leads to the following:

\begin{corollary}
\label{cor:generalize}

Suppose $\XX$ and $\YY$ are matrices with the same null space
such that $\XX \preceq \YY$,
$\bb = \YY \xxbar$, and $\xx$ is an arbitrary vector.
If $\matz = \textsc{Sample} (\maty_1 \ldots \maty_m, X, \er_1 \ldots \er_m, \frac{1}{10})$,
and $\xx'$ generated using
\[
\xx' = \vecx - \frac{1}{10} \ZZ^{\dag} \left( \YY \xx - \bb \right).
\]
Then
\begin{align*}
\expct{r, i_1, i_2, \ldots i_r}{\nbr{\bar{\vecx}
- \left( \vecx - \frac{1}{10} \matz^{\dag} \left( \maty \vecx - \vecb \right) \right)}_{\maty}}
\leq \left( 1 - \frac{1}{80} \right) \nbr{\bar{\vecx} - \vecx}_{\maty}
\end{align*}
\end{corollary}

\section{Expected Inverse Moments}

\label{sec:moments}
We now prove the bounds on $\ZZ^{-1}$ and $\ZZ^{-1} \YY \ZZ^{-1}$
stated in Lemma~\ref{lem:moments}.
For simplicity, we define $\vecnv_j:=\maty^{\frac{-1}{2}}\vecv_j$, and $\matnx:=\maty^{\frac{-1}{2}}X\maty^{\frac{-1}{2}}$.  Note that, $\sum_{j = 1}^m \vecnv_j\vecnv_j^T = \mati$, while
\begin{align*}
\vecnv_i^T \matnx^{-1} \vecnv_i 
&= \vecv_i^T \matx^{-1} \vecv_i  \\
&= \trace{\matx^{-1} \vecv_i \vecv_i^T}\\
&=\er_i.
\end{align*}

The following lemma is then equivalent to Lemma~\ref{lem:moments}.  
\begin{lemma}
\label{lem:moments1}
Suppose $\matny_i = \vecnv_i \vecnv_i^T$ are rank one matrices with $\sum_{j = 1}^m \vecnv_j\vecnv_j^T = \mati$,
$\matnx$ is a positive definite matrix satisfying $\matnx \preceq \mati$ and $\er_1 \ldots \er_m$ are values that satisfy
$\er_i \geq \trace{\matnx^{-1} \matny_i}$,
and $0 < \delta < 1$ is an arbitrary parameter.
Then the matrix $\matnz = \textsc{Sample}(\matny_1 \ldots \matny_m, \matnx,
\er_1 \ldots \er_m, \delta)$ satisfies:
\begin{enumerate}
\item $\expct{r, i_1 \ldots i_r}{\vecx^T \matnz^{-1} \vecx} \geq \frac{1}{3} \vecx^T \matni \vecx$, and
\item $\expct{r, i_1 \ldots i_r}{\vecx^T \matnz^{-1} \vecx} \leq \frac{1}{1 - 2 \delta} \vecx^T\matni\vecx $, and
\item $\expct{r, i_1 \ldots i_r}{\vecx^T \matnz^{-2} \vecx} \leq \frac{1}{1 - 3 \delta} \vecx^T \matni \vecx$.
\end{enumerate}
\end{lemma}
In remainder of this section, we prove the above lemma. To analyze the \textsc{Sample} algorithm, it will be helpful to keep track of its intermediate steps. Hence, we define $\matnz_0$ to be the initial value of the sample sum matrix $\matnz$.
This corresponds to the initial value of $Z$ from Line~\ref{line:initZ} in the pseudocode of Figure~\ref{fig:sample}, and  $\matnz_0 = S$. We define $\matnz_j$ to be the value of $\matnz$ after $j$ samples. Thus $\matnz_{j+1} = \matnz_j+\frac{\delta}{\er_{i_{j+1}}}\vecnv_{i_{j+1}}\vecnv_{i_{j+1}}^T$ where $i_{j+1}$ is chosen with probability proportional to $\er_{j+1}$.

Throughout this section, we use $\delta$ to refer to the constant as defined in lemma~\ref{lem:moments1} and let
$$t := \delta^{-1}\sum_{i = 1}^m{\er_i}.$$

The following easily verifiable fact will be useful in our proofs.
 \begin{fact}
\label{fact:oneSampleAvg}
With variables as defined in lemma~\ref{lem:moments1}, each sample $\frac{\delta}{\er_{i_j}}\vecnv_{i_j}\vecnv_{i_j}^T$ obeys
$$\expct{i_j}{\frac{\delta}{\er_{i_j}}\vecnv_{i_j}\vecnv_{i_j}^T} = \frac{1}{t} \mati$$
\end{fact}

As we will often prove spectral bounds on the inverse of matrices,
the following simple statement about positive definite matrices
is very useful to us.
 \begin{fact}
\label{fact:invOpMonotone}
Given positive definite matrices $A$ and $B$ where $A \preceq B$,
$$ B^{-1} \preceq A^{-1}. $$
\end{fact}

The lower bound on $\WW^{-1}$ can be proven using these two facts,
and a generalization of the arithmetic mean (AM)
- harmonic mean (HM) inequality for matrices
by Sagae and Tanabe~\cite{SagaeT94}.

\begin{lemma}[matrix AM-HM inequality, part of Theorem 1 of ~\cite{SagaeT94}]
\label{lem:amhm}
If $\ww_1,  \ldots, \ww_r$ are positive numbers such that $\ww_1 + \ldots + \ww_r = 1$, and let $\MM_1, \ldots, \MM_r$ be positive definite matrices.
Then
\[
\left( \ww_1 \MM_1^{-1} + \ldots + \ww_r \MM_r^{-1}  \right)^{-1}
\preceq \ww_1 \MM_1+ \ldots + \ww_r \MM_r.
\]
\end{lemma}

\begin{Proofof}{Lemma~\ref{lem:moments1}, Part~\ref{part:firstLower}}
For all $j$, the matrix $\vecnv_{j}\vecnv_{j}^T$ is positive semidefinite. Hence, using the fact~\ref{fact:oneSampleAvg},
\begin{align*}
\expct{r,i_1,...,i_{r}}{\matnz}
& \preceq \expct{i_1,\ldots,i_{r}}{\matnz \vert r = 2t } \\
& = {S + \sum_{j=1}^{2t}\expct{i_j}{\frac{\delta}{\er_{i_j}}\vecnv_{i_j}\vecnv_{i_j}^T}}  \preceq 3 \mati
\end{align*}
Consequently, by the AM-HM bound from Lemma~\ref{lem:amhm} gives
\[
\expct{r,i_1,...,i_{r}}{ \matnz^{-1} }^{-1} \preceq \left(3 \mati\right)^{-1}.
\]
Inverting both sides using Fact~\ref{fact:invOpMonotone} gives the result.
\QED
\end{Proofof}

We can now focus on proving the two upper bounds.
One of the key concepts in our analysis is the harmonic sum,
named after the harmonic mean,
\begin{align}
\hsum{x}{y} & \defeq \frac{1}{1/x + 1/y}.
\end{align}
The following property of the harmonic sum plays a crucial role in our proof:
 \begin{fact}
\label{fact:hSumExpct}
If $X$ is a positive random variable and $\alpha > 0$ is a constant, then
\begin{align*}
\expct{}{\hsum{X}{\alpha}} & \leq\hsum{\expct{}{X}}{\alpha}.
\end{align*}
\end{fact}

\Proof
Follows from Jensen's inequality since
\begin{align*}
\hsum{X}{\alpha} =\frac{1}{\frac{1}{X}+\frac{1}{\alpha}}=\alpha\left(1-\frac{\alpha}{X+\alpha}\right)
\end{align*}
 is a concave function in $\alpha$ when $\alpha > 0$.
\QED

We will also use a matrix version of this:
\begin{fact}
\label{fact:harmonicMatrix}
For any unit vector $\vecv$,
positive definite matrix $\mata$, and scalar $\alpha > 0$
\begin{align*}
\vecv^T \left(\mata + \alpha \mati \right)^{-1} \vecv
& \leq \hsum{\vecv^T \mata^{-1} \vecv}{1/\alpha}
\end{align*}
\end{fact}

\Proof
By a change of basis if necessary, we can assume $\mata$ is
a diagonal matrix with positive entries $\left(a_{1},...,a_{n}\right)$ on its
diagonal. Then $\vecv^{T}\left(\mata+\alpha I\right)^{-1}\vecv=\sum_{i=1}^{n}\frac{v_{i}^{2}}{a_{i}+\alpha}=\mathbb{E}\left[\text{HrmSum}\left(X,\alpha\right)\right]$
where $X$ is a random variable which satisfying $X=\frac{1}{a_{i}}$
with probability $\vecv_{i}^{2}$. Then by Fact~\ref{fact:hSumExpct} we have $$
\sum_{i=1}^{n}\frac{\vecv_{i}^{2}}{a_{i}+\alpha}=\mathbb{E}\left[\text{HrmSum}\left(X,\frac{1}{\alpha}\right)\right]\leq\text{HrmSum}\left(\mathbb{E}\left[X\right],\frac{1}{\alpha}\right)=\text{HrmSum}\left(\vecv^{T}\mata^{-1}\vecv,\frac{1}{\alpha}\right)$$
because $\mathbb{E}\left[X\right]=\sum_{i}\frac{\vecv_{i}^{2}}{a_{i}}=\vecv^{T}\mata^{-1}\vecv$.
\QED

\begin{fact}
\label{fact:convexf}
The function $\func{\mathh,\vecv}(x)$ defined by
$$\func{\mathh,\vecv}(x):=\vecv^{T}\left(\mathh+\frac{x}{t}I\right)^{-1}\vecv$$
is convex in $x$ for any fixed choices of
vector $\vecv$ and positive definite matrix $\mathh$ . 
\end{fact}

\Proof
By a change of basis, we can assume $\mathh$ to be diagonal matrix
without loss of generality.
Let its diagonal entries be $\left(a_{1},...,a_{n}\right)$.
Since $\HH$ is positive definite, $a_{i} > 0$.
The result then follows from
$$ \func{\mathh,\vecv}(x) = \sum_{i} \frac{v_{i}^{2}}{a_{i}+\frac{x}{t}} $$
which is a convex function in $x$.
\QED

This implies that 
\begin{align*}
\vecv^T \matnz_i^{-1} \vecv + \func{\matnz_j,\vecv}'(0)
& = \func{\matnz_j,\vecv}(0) + (1 - 0) \func{\matnz_j,\vecv}'(0)\\
& \leq \func{\matnz_j,\vecv}(1).
\end{align*}
Also, when $\vecv$ is a unit vector, we have by Fact~\ref{fact:harmonicMatrix}:
\begin{align*}
\func{\matnz_j,\vecv}(1) \leq \operatorname{HrmSum}(\vecv^T \matnz_j^{-1} \vecv, t),
\end{align*}
which rearranges to
\begin{align*}
\func{\matnz_j,\vecv}'(0)
& \leq \operatorname{HrmSum}(\vecv^T \matnz_j^{-1} \vecv, t) - \vecv^T \matnz_j^{-1}\vecv.
\end{align*}
Also, note that:
\begin{align*}
\func{\matnz_j,\vecv}'(x)
& =  -\frac{1}{t} \vecv^T (\matnz_j + (x/t) I)^{-2} \vecv\\
\func{\matnz_j,\vecv}'(0) 
& = -\frac{1}{t} \vecv^T \matnz_j^{-2} \vecv.
\end{align*}
So
\begin{align}
 -\frac{1}{t} \vecv^T \matnz_j^{-2} \vecv
\leq
\operatorname{HrmSum}(\vecv^T \matnz_j^{-1} \vecv, t) - \vecv^T \matnz_j^{-1}\vecv.
\label{eq:derivf}
\end{align}

We can also obtain a spectral lower bound $\matnz_{j + 1}^{-1}$
in terms of $\matnz_{j}^{-1}$ and $\matnz_{j}^{-2}$.
using the Sherman-Morrison formula.

\begin{lemma}
\label{lemma:expIncrease}
$\cexpct{{i_{j+1}}}{\matnz_{j+1}^{-1}}{\matnz_{j}} \preceq \matnz_{j}^{-1}-\frac{\left(1-\delta\right)}{t}\matnz_{j}^{-2}$

\end{lemma}

\Proof

The Sherman-Morrison formula says that adding a single sample $\vecnnv_{j}\vecnnv_{j}^{T}:=\frac{\delta}{\er_{i_{j+1}}}\vecnv_{i_{j+1}}\vecnv_{i_{j+1}}^{T}$ to $\matnz_j$ gives: 
$$ (\matnz_j + \vecnnv_{j}\vecnnv_{j}^T)^{-1} = \matnz_j^{-1} - \frac{\matnz_j^{-1} \vecnnv_{j}\vecnnv_{j}^{T} \matnz_j^{-1}}{1 + \vecnnv_{j}^T \matnz_j^{-1} \vecnnv_{j}}.  $$

We then have 
\begin{align*}
(\matnz_j + \vecnnv_{j}\vecnnv_{j}^T)^{-1}
&= \matnz_j^{-1} - \frac{\matnz_j^{-1} \vecnnv_{j}\vecnnv_{j}^{T} \matnz_j^{-1} }{1 + \vecnnv_{j}^T \matnz_j^{-1}\vecnnv_{j}} \\
&\preceq \matnz_j^{-1} - \frac{\matnz_j^{-1} \vecnnv_{j}\vecnnv_{j}^{T} \matnz_j^{-1} }{1 + \delta}\\
&\preceq \matnz_j^{-1} - (1 - \delta)\matnz_j^{-1} \vecnnv_{j}\vecnnv_{j}^{T} \matnz_j^{-1}. 
\end{align*}

Hence, 
\begin{align*}
\cexpct{{i_j}}{(\matnz_j + \vecnnv_{j}\vecnnv_{j}^T)^{-1}}{\matnz_j}
&\preceq \matnz_j^{-1} - (1 - \delta)\matnz_j^{-1} \expct{i_j}{\vecnnv_{j}\vecnnv_{j}^{T}} \matnz_j^{-1} \\
&\preceq \matnz_j^{-1} - \frac{\left(1-\delta\right)}{t}\matnz_j^{-1}\mati\matnz_j^{-1} \\
&\preceq \matnz_{j}^{-1}-\frac{\left(1-\delta\right)}{t}\matnz_{j}^{-2}.
\end{align*}
\QED

Combining these two bounds leads us to an upper bound
for $\expct{}{\matnz_j^{-1}}$.

\begin{Proofof}{Lemma~\ref{lem:moments1}, Part~\ref{part:firstUpper}}
Combining Lemma~\ref{lemma:expIncrease} and Equation~\ref{eq:derivf}, we have 
\begin{align*}
\vecv^T\cexpct{{i_{j+1}}}{(\matnz_j + \vecnnv_{j+1}\vecnnv_{j+1}^T)^{-1}}{\matnz_j}\vecv
& \leq \vecv^T \matnz_j^{-1} \vecv - \frac{1-\delta}{t} \vecv^T \matnz_j^{-2} \vecv\\
&\leq \vecv^T \matnz_j^{-1} \vecv - (1-\delta) \left(\vecv^T \matnz_j^{-1} \vecv - \operatorname{HrmSum}\left(\vecv^T \matnz_j^{-1} \vecv, t\right)\right)\\
& = \delta \vecv^T \matnz_j^{-1} \vecv + (1 - \delta) \operatorname{HrmSum}\left(\vecv^T \matnz_j^{-1} \vecv, t\right)
\end{align*}

If we now include the choice of $\matnz_j$ in the expectation:
\begin{align*}
\expct{{i_1},...,{i_{j+1}}}{\vecv^T \matnz_{j+1}^{-1} \vecv}
&\leq \expct{{i_1},...,{i_j}}{ \delta \vecv^T \matnz_j^{-1} \vecv +
 (1-\delta) \operatorname{HrmSum}\left(\vecv^T \matnz_j^{-1} \vecv, t\right)} \\
& = \delta \expct{{i_1},...,{i_{j}}}{ \vecv^T \matnz_j^{-1} \vecv}
 + (1-\delta) \expct{{i_1},...,{i_j}}{\operatorname{HrmSum}\left(\vecv^T \matnz_j^{-1} \vecv, t \right)}.
\end{align*}
Applying Fact~\ref{fact:hSumExpct} with $X = \vecv^T \matnz_j^{-1} \vecv$
and $a = t$ gives
\begin{align}
\expct{{i_1},...,{i_{j+1}}}{\vecv^T \matnz_{j+1}^{-1} \vecv} \leq \delta \expct{{i_1},...,{i_j}}{ \vecv^T \matnz_j^{-1} \vecv}
 + (1-\delta) \operatorname{HrmSum} \left(\expct{i_1,...,i_j}{\vecv^T \matnz_j^{-1} \vecv}, t\right).
 \label{eq:iterExpct}
\end{align}

For convenience, we define $E_j:= \expct{{i_1},...,{i_j}}{\vecv^T \matnz_{j}^{-1} \vecv}$.
So inequality \ref{eq:iterExpct}  can be written as
\begin{equation*}
E_{i + 1}
\leq \delta E_{i}  + \left( 1 - \delta \right) \operatorname{HrmSum}\left(E_i, t\right)
\end{equation*}
Also, since we start with $\matnz_0 = \maty^{\frac{-1}{2}}\matx\maty^{\frac{-1}{2}}$, we have $\matnz_j \succeq \maty^{\frac{-1}{2}}\matx\maty^{\frac{-1}{2}}$. Thus, by fact~\ref{fact:invOpMonotone}  
$$\matnz_j^{-1} \preceq (\maty^{\frac{-1}{2}}\matx\maty^{\frac{-1}{2}})^{-1} = \maty^{\frac{1}{2}}\matx^{-1}\maty^{\frac{1}{2}}.$$
So $\trace{\matnz_j^{-1}} \leq \trace{\maty^{\frac{1}{2}}\matx^{-1}\maty^{\frac{1}{2}}} \leq \sum_{i=1}^m \er_i = t\delta$, 
and we have $\vecv^T \matnz_j^{-1} \vecv \leq \nbr{\matnz_j^{-1}} \leq t \delta$,
so $E_j \leq t \delta < t$.
This lets us write:
\begin{align*}
E_{j + 1}
& = \delta E_j + \frac{1 - \delta}{\frac{1}{E_j} + \frac{1}{t}}\\
& = \frac{1 + \frac{\delta E_j }{ t }}{\frac{1}{E_j} + \frac{1}{t}}\\
& \leq \frac{1}{\left(\frac{1}{E_j}
	+ \frac{1}{t}\right) \left( 1 - \frac{\delta E_j}{t} \right)}\\
& = \frac{1}{\frac{1}{E_j} + \frac{1}{t} - \frac{\delta}{t} - \frac{\delta E_j}{t^2}}\\
&\leq \frac{1}{1/E_j + (1-2 \delta)/t}.
\end{align*}
So
\begin{align*}
\frac{1}{E_{j + 1}}
\geq & \frac{1}{E_j} + \left( 1 - 2\delta \right)/t
\end{align*}
Then it follows by induction that after $t$ steps $$\frac{1}{E_j} \geq \left( 1 - 2\delta \right).$$ Thus we have proved
\begin{align}
\expct{{i_1},...,{i_t}}{\vecv^T \matnz_t^{-1} \vecv}\leq \frac{1}{1-2\delta}.
\label{eq:inverseUpperBoundAtT}
\end{align}
 Additionally, for any integer $r \geq t$, $\matnz_r \succeq \matnz_t$, so fact~\ref{fact:invOpMonotone} gives $\matnz_r^{-1} \preceq \matnz_t^{-1}$. This means that with $r$ chosen uniformly at random in the interval
$[t,2t-1]$, we have
 $$\expct{r,{i_1},...,{i_r}}{\vecv^T \matnz_r^{-1} \vecv}\leq \frac{1}{1-2\delta}.$$ 
 \QED
 \end{Proofof}
  
It remains to upper bound $\matnz_r^{-2}$.
Here we use the same proof technique in reverse, by showing that the
increase in $\matnz_r^{-1}$ is related to $\matnz_r^{-2}$.
Lemma~\ref{lem:moments1}, Part~\ref{part:firstUpper} gives that
the total increase between $t$ and $2 t - 1$ is not too big.
Combining this with the fact that we chose $r$ randomly gives that the
expected increase at each step, and in turn the expected value of
$\matnz_r^{-2}$ is not too big as well.

\begin{Proofof}{Lemma~\ref{lem:moments1}, Part~\ref{part:secondUpper}}
Recall that the expected value of $\vecv^T \matnz_{j+1}^{-1} \vecv - \vecv^T \matnz_j^{-1} \vecv$,
 conditional on $\matnz_j$, was at most
\begin{align*}
\vecv^T \matnz_{j+1}^{-1} \vecv - \vecv^T \matnz_j^{-1} \vecv
\leq & (1-\delta) f'(0) = \frac{-(1-\delta)}{t} \vecv^T W_j^{-2} \vecv
\end{align*}
Taking expectation over everything gives:
\begin{align*}
\expct{i_1,\ldots,i_{j+1}}{\vecv^T \matnz_{j+1}^{-1} \vecv} - \expct{i_1,\ldots,i_{j}}{\vecv^T \matnz_j^{-1} \vecv}
\leq
\expct{i_1,\ldots,i_{j}}{\frac{-(1-\delta)}{t} \vecv^T \matnz_j^{-2} \vecv}
\end{align*}
Telescoping this gives
\begin{align*}
\expct{i_1,\ldots,i_{2t}}{\vecv^T \matnz_{2t-1}^{-1} \vecv} - \expct{i_1,\ldots,i_t}{{\vecv^T \matnz_t}^{-1} \vecv}
&\leq \sum_{j = t}^{2t-1} \expct{i_1,\ldots,i_{j}}{\frac{-(1-\delta)}{t} \vecv^T \matnz_j^{-2} \vecv}\\
\frac{1}{t} \sum_{j = t}^{2t-1} \expct{i_1,\ldots,i_j}{\vecv^T \matnz_j^{-2} \vecv}
&\leq \frac{1}{1 - \delta} \expct{i_1,\ldots,i_t}{\vecv^T \matnz_t^{-1} \vecv} \leq  \frac{1}{(1-2\delta)(1-\delta)},
\end{align*}
where the last inequality follows from equation~\ref{eq:inverseUpperBoundAtT}. This implies that for an integer $r$ chosen uniformly at random in the interval $[t,2t-1]$, we have
$$\expct{r,{i_1},...,{i_r}}{\vecv^T \matnz_r^{-2} \vecv} \leq \frac{1}{(1-2\delta)(1-\delta)} < \frac{1}{1-3\delta}.$$
\QED
\end{Proofof}

\section{Application to Solving SDD linear systems}
\label{sec:sdd}

We now describe a faster algorithm for solving SDD linear systems
that relies on preconditioners that make progress in expectation.
The reduction from solving these systems to solving graph Laplacians
of doubled size was first shown by Gremban and Miller~\cite{Gremban96:thesis}.
This reduction is also well-understood for approximate 
solvers~\cite{SpielmanTengSolver}, and in the presence of fixed
point round-off errors~\cite{KelnerOSZ13}.
As a result, we only address solving graph Laplacians in our presentation.

The Laplacian of a weighted graph $G$ is an $n \times n$
matrix containing the negated weights in the off-diagonal entries
and weighted degrees in the diagonal entries:
\begin{definition}
\label{def:graphLaplacian}
The graph Laplacian $\laplacian_G$ of a weighted graph $G = (V, E, \ww)$
with $n$ vertices is an $n \times n$ matrix whose entries are:
\begin{align*}
\laplacian_{G,uv} &=
\begin{cases}
\sum_{v \neq u} \ww_{uv} &~~~\text{if }u = v,\\
-\ww_{uv} &~~~\text{otherwise.}
\end{cases}
\end{align*}
\end{definition}

The recursive preconditioning framework due to Spielman and Teng
extends the ideas pioneered by Vaidya~\cite{Vaidya91}.
It generates graph preconditioners, called \emph{ultra-sparsifiers},
by sampling a number of edges to supplement a carefully chosen spanning tree.
Using the notation introduced in Section~\ref{sec:overview}, this corresponds
to setting $\XX$ to the graph Laplacian of the tree and the $\YY_i$s to
the graph Laplacians of the off-tree edges.

The key connection between the statistical leverage score of a tree
and combinatorial stretch of an edge was observed by
Spielman and Woo~\cite{SpielmanW09}.
\begin{fact}
\label{fact:stretch}
The statistical leverage score of the rank-1 matrix corresponding to
an edge w.r.t. a tree is equal to its combinatorial stretch w.r.t. that tree.
\end{fact}

The reason that it is crucial to pick $\XX$ to be a tree is that
then the sizes of the recursive subproblems only depend on the
number of $\YY_i$'s considered within.
Similar to previous solvers, our algorithm is recursive.
However, it chooses a different graph at each iteration, so that many distinct graphs are given in calls at the
same level of the recursion.
As a result, we will define an abstract Laplacian solver
routine for our analyses.

\begin{definition}
\label{def:lapSolver}
A routine $\textsc{Solver}(\cdot)$ is said to be a \emph{Laplacian solver} when it takes as input a tuple $(G, T,\er,\bb, \epsilon)$, where $G$ is a graph, $T$ a spanning tree of this graph, and $\er$ upper bounds on the combinatorial stretch of the off-tree edges of $G$ wrt. $T$, and the routine returns as output a vector $\xx$ such that
\[
\norm{\xx - \LL_{G}^{\dag} \bb}_{\LL_{G}} \leq \epsilon \norm{\LL_{G}^{\dag} \bb}_{\LL_{G}}.
\]
\end{definition}

The following lemma about size reduction can be derived from
partial Cholesky factorization.
A detailed proof of it can be found in Appendix C of~\cite{Peng:thesis}.

\begin{restatable}{lemma}{greedyElimination}
\label{lem:greedyElimination}
Given a graph-tree tuple $(H, T, \er)$ with $n$ vertices
and $m'$ off-tree edges, and
and a Laplacian solver $\textsc{Solver}$, there is a routine
$\textsc{Eliminate\&Solve}(H,T,\er, \textsc{Solver}, \vecb, \epsilon)$ that
for any input $\vecb = \laplacian_H \bar{\vecx}$, performs $\Oh(n + m')$ operations
plus one call to $\textsc{Solver}$ with a graph-tree tuple $(H', T', \er')$ with $\Oh(m')$ vertices and edges, the same bounds for the stretch of off-tree edges,
and accuracy $\epsilon$ and returns a vector $\vecx$ such that
\begin{align*}
\nbr{\bar{\vecx} - \vecx}_{\laplacian_H} \leq \epsilon \nbr{\bar{\vecx}}_{\laplacian_H}.
\end{align*}
\end{restatable}

With this in mind, one way to view the recursive preconditioning
framework is that it gradually reduces the number of edges using the
statistical leverage scores obtained from a tree.
For this, Koutis et al.~\cite{KoutisMP11} used the low-stretch
spanning tree algorithms~\cite{AlonKPW95,ElkinEST08,AbrahamBN08,AbrahamN12}.
However, the state of art result due to Abraham and 
embeddings takes $\Oh(m \log{n} \log\log{n})$ time to construct.

Instead, we will use the low-stretch embeddings given by Cohen et al.~\cite{CohenMPPX13}.
Their result can be summarized as follows:

\begin{lemma}
\label{lem:hasTree}
Given a graph $\hat{G}$ with $n$ vertices, $m$ edges, and any constant $0 < p < 1$,
we can construct in $\Oh(m\log\log{n}\log\log\log{n})$ time in the RAM model
a graph-tree tuple $(G, T, \er)$ and associated bounds on stretches of edges $\er$ such that
\begin{enumerate}
\item $G$ has at most $2n$ vertices and $n + m$ edges, and
\item $\nbr{\er}_p^{p} \leq \Oh(m \log^{p}n)$, and
\item
\label{item:goodApprox}
there is a $|V_{\hat{G}}| \times |V_{G}|$ matrix $\PPi$
with one $1$ in each row and zeros everywhere else such that:
\[
\frac{1}{2} \LL_{\hat{G}}^{\dag}
\preceq \PPi_1 \PPi \LL_{G}^{\dag}  \PPi^T \PPi_1^T
\preceq \LL_{\hat{G}}^{\dag}.
\]
Note that $\PPi$ maps some vertices of $G$ to unique vertices of $\hat{G}$,
and $\PPi^T$ maps each vertex of $\hat{G}$ to a unique vertex in $G$.
\end{enumerate}
\end{lemma}

The spectral guarantees given in Part~\ref{item:goodApprox}
allow the solver for $\LL_{G}$ to be converted
to a solver for $\LL_{\hat{G}}$ while preserving the error quality.

\begin{restatable}{fact}{errorTransfer}
\label{fact:errorTransfer}
Let $\PPi$ and $\PPi_1$ be the two projection matrices defined
in Lemma~\ref{lem:hasTree} Part~\ref{item:goodApprox}.
For a vector $\bbhat$, if $\xx$ is a vector such that
\[
\norm{\xx - \LL_{G}^{\dag} \PPi^T \PPi_1^T \bbhat}_{\LL_{G}}
\leq \epsilon \norm{\LL_{G}^{\dag} \Pi^T \PPi_1^T \bbhat}_{\LL_G},
\]
for some $\epsilon > 0$.
Then the vector $\xxhat = \PPi_1 \PPi\xx$ satisfies
\[
\norm{\xxhat - \PPi_1 \PPi \LL_{G}^{\dag}  \PPi^T \PPi_1^T \bbhat}
_{\left(\PPi_1 \PPi \LL_{G}^{\dag}  \PPi^T \PPi_1^T\right)^{\dag}}
\leq \epsilon \norm{\PPi_1 \PPi \LL_{G}^{\dag}  \PPi^T \PPi_1^T\bbhat}
_{ \left(\PPi_1 \PPi \LL_{G}^{\dag}  \PPi^T \PPi_1^T\right)^{\dag}}.
\]
\end{restatable}

Therefore, a good solution to $\LL_{G} \xx = \PPi^T \widehat{\bb}$
also leads to a good solution to $\LL_{\widehat{G}} \widehat{\xx} = \widehat{\bb}$.
The constant relative error can in turn be corrected using preconditioned Richardson
iteration described in Section~\ref{sec:overview}.
For the rest of our presentation, we will focus on solving linear systems
in settings where we know small bounds to $\nbr{\er}_p^{p}$.

As \textsc{Sample} will sample edges with high stretch, as well as
tree edges, we need to modify its construction  bounding both the
number of off-tree edges, and the total off-tree $\ell_p$-stretch.
Pseudocode of this modified algorithm for generating a preconditioner
is given in Figure~\ref{fig:randPrecon}.

\begin{figure}
\vskip 0.2in
\centering
\fbox{
\begin{minipage}{6in}
\noindent $(H, T') = \textsc{RandPrecon} (G, T, \er, \delta)$,
where $G$ is a graph, $T$ is a tree, $\er$ are upper bounds of
the stretches of edges in $G$ w.r.t. $T$,
and $\delta < 1$ is an arbitrary parameter.
\begin{enumerate}
\item Let $\matx = \laplacian_T$, $\maty = \laplacian_G$,
$\maty_i$ be the rank-1 matrix corresponding to each edge.
\item Set $\hat{\er}$ to be the same as $\er$ for non tree-edges,
and $1$ for all tree edges.
\item Repeat
\begin{enumerate}
\item $\matz = \textsc{Sample}\left(\maty, \matx, \hat{\er}, \frac{1}{10} \right)$.
\item Set
\begin{enumerate}
\item $H$ be the edges corresponding to $\matz$, and
\item $T'$ be the edges corresponding to the combinatorial components in $T$, and
\item $\er'$ to be $\delta$ times the number of times each off-tree edge is sample.
\end{enumerate}
\end{enumerate}
\item Until the number of off-tree edges in $H$ is at most $4800 \nbr{\er}_p^{p}$,
and $\nbr{\er'}_p^{p} \leq 480 \nbr{\er}_p^{p}$.
\item Return $(H, T', \er')$.
\end{enumerate}
\end{minipage}
}
\caption{Generation of a Randomized Preconditioner}
\label{fig:randPrecon}
\end{figure}

We start by proving some crude guarantees of this algorithm.

\begin{lemma}
\label{lem:randPrecon}
$\textsc{RandPrecon}(G, T, \er, \frac{1}{10})$ runs in expected $\Oh(m + \nbr{\er}_p^{p})$ time
and produces a graph-tree tuple $(H, T, \er')$ such that
\begin{enumerate}
\item the number of off-tree edges in $H$ is at most $\Oh( \nbr{\er}_p^{p} )$, and
\item $\nbr{\er'}_p^{p} \leq \Oh( \nbr{\er}_p^{p} )$, and
\item for any pair of vectors $\vecx$
and $\vecb = \laplacian_G \bar{\vecx}$, we have
\begin{align*}
\expct{H}{\nbr{\bar{\vecx}
- \left( \vecx - \frac{1}{10} \laplacian_H^{\dag} \left( \laplacian_G \vecx - \vecb \right) \right)}_{\laplacian_G}}
\leq \left( 1 - \frac{1}{160} \right) \nbr{\bar{\vecx} - \vecx}_{\laplacian_G}
\end{align*}
\end{enumerate}
\end{lemma}

\Proof
For an edge $e$, let $X_e$ be a random variable indicating the
number of times that $e$ is sampled.
The call to \textsc{Sample} samples $\frac{s}{\delta}$ edges
where  $s$ is the total stretch of all edges.
In each of these iterations, $e$ is sampled with
probability $\frac{\er_e}{s}$ where $\er_e = \trace{\vecindicator_e^T \laplacian_T^{\dag} \vecindicator_e}$.
This means the expected value of $X_e$ is
\begin{align*}
\expct{}{X_e}
& \leq \frac{3}{2}\frac{s}{\delta} \frac{\er_e}{s}\\
& = \frac{3}{2}\delta^{-1} \er_e
= 15 \er_e.
\end{align*}
For an edge $e$, if $\er_e \geq 1$, then $\er^{p}_e \geq 1$;
otherwise, $\er_e \leq\er^{p}_e$.
Therefore we have that the expected number of distinct edges
added to the tree is less than $15 \nbr{\er}_p^{p}$.
Markov's inequality then gives that we sample more than
$4800 \nbr{\er}_p^{p}$ edges with probability
at most $\frac{1}{320}$.

For the expected stretch, note that as $T$ is added to $H$,
the stretch of an edge can only decrease.
Combined with the fact that each sampled edge has stretch $\delta$ with respect to $T$:
\begin{align*}
\expct{}{(\er'_e)^p}
\leq \expct{}{\left( \delta X_e\right)^{p} }
\leq \expct{}{\delta X_e}^{p}
& \leq \frac{3}{2}\er_e^{p}.
\end{align*}
So the expected total $\ell_p$-stretch of all off-tree edges
is at most $\frac{3}{2}\er_e^{p}$.
Applying Markov's inequality once again gives that the
probability of $\nbr{\er'}_p^{p} \leq 480 \nbr{\er}_p^{p}$
is also at most $\frac{1}{320}$.

Taking a union bound gives that each sample $H$ fails the conditions
with probability at most $\frac{1}{160}$.
This means that the loop is expected to terminate in $\Oh(1)$ iterations.
Also, this means that the expected deviation in $H$ only increases
by a constant factor, giving
\begin{align*}
\expct{H}{\nbr{\bar{\vecx}
- \left( \vecx - \frac{1}{10} \laplacian_H^{\dag} \left( \laplacian_G \vecx - \vecb \right) \right)}_{\laplacian_G}}
& \leq \frac{1}{1 - \frac{1}{160}} \left( 1 - \frac{1}{80} \right) \nbr{\bar{\vecx} - \vecx}_{\laplacian_G}\\
& \leq \left(1 - \frac{1}{160} \right)  \nbr{\bar{\vecx} - \vecx}_{\laplacian_G}.
\end{align*}
\QED

We can then apply the elimination routine from
Lemma~\ref{lem:greedyElimination} to obtain
a high-quality solution to a linear system by solving a small number
of systems whose edge count is $\Oh(\nbr{\er}_p^{p})$.

However, note that the error is in the $\laplacian_H$-norm.
To relate this to the $\laplacian_G$-norm, we can use a spectral bound
derived from matrix concentration bounds.
Such a bound is central in operator based solvers by Koutis
et al.~\cite{KoutisMP10,KoutisMP11}, while we feel our use
of it here is more tangential.

\begin{lemma}
\label{lem:spectralBound}
There exists a constant $c$ such that for any graph-tree tuple
$G$, $T$, $\er$, $H = \textsc{RandPrecon}(G, T, \er, \frac{1}{10})$ satisfies
\begin{align*}
\frac{1}{c \log{n}} \laplacian_G
&\preceq \laplacian_H \preceq c \log{n} \laplacian_G
\end{align*}
 with high probability.
\end{lemma}

We prove this bound in Appendix~\ref{sec:whp}.
It means that the decrease in energy can still be guaranteed if
we set $\epsilon = \Oh(\frac{1}{c_s \log{n}} )$ in our bounds.
We can also check whether we have reached such an error using
coarser solvers.

\begin{lemma}
\label{lem:coarseSolver}
There exist a constant $c_Z$ such that given a graph-tree
tuple $G$, $T$, $\er$, we can construct with high probability
a linear operator $\matz$ such that under exact arithmetic
\begin{enumerate}
\item $\laplacian_G^{\dag} \preceq \matz \preceq c_Z \log^{4}n \laplacian_G^{\dag}$, and
\item given any vector $\vecb$, $\matz \vecb$ can be evaluated in
$\Oh(m + \nbr{\er}_p^{p})$ time where $p$ is any constant $> 1/2$.
\end{enumerate}
\end{lemma}

\Proof
Consider scaling up the tree by a factor of $\log^2{n}$ and scaling
down the bounds on leverage scores accordingly to obtain $G', T', \er'$.
Then $\LL_G \preceq \LL_{G'} \preceq \log^2{n} \LL_G$
and Lemma~\ref{lem:randPrecon} gives that $H = \textsc{randPrecon}(G', T', \er')$ has $\Oh(\nbr{\er'}_p^{p}) = \Oh(\log^{-2p}n \nbr{\er}_p^{p})$ off-tree edges, and
\[
\frac{1}{c \log{n}} \LL_{G'} \preceq \LL_H \preceq c \log{n} \LL_{G'}.
\]

Applying partial Cholesky factorization on $H$ and then the solver
algorithm by Koutis et al.~\cite{KoutisMP11} then gives an operator $\ZZ$
such that
\begin{align*}
\frac{1}{2} \LL_H^{\dag} \preceq \ZZ \preceq 2 \LL_H^{\dag},
\end{align*}
and $\ZZ \bb$ can be evaluated in
$\Oh(m + \log^{-2p}n \nbr{\er}_p^{p} \log{n} \log\log^{1/2}n )
\leq O(m + \nbr{\er}_p^{p})$ time.
Propagating the error guarantees then gives
\[
\ZZ \preceq 2 \LL_H^{\dag} \preceq 2 c\log{n} \LL_{G'}^{\dag} \preceq 2 c \log{n} \LL_{G}^{\dag},
\]
for the upper bound, and
\[
\ZZ \succeq \frac{1}{2} \LL_H^{\dag}
\succeq \frac{1}{2 c \log{n}} \LL_{G'}^{\dag}
\succeq \frac{1}{2 c \log^{3}{n} } \LL_G^{\dag},
\]
for the lower bound.
Scaling $\ZZ$ by a factor of $2 c \log^{3}n$ then gives the required operator.
\QED

Using this routine allows us to  convert the expected convergence
to one that involves expected running time, but converges with high probability.
This is mostly to simplify our presentation and we believe such
a dependency can be removed.
Using this routine leads us to our randomized preconditioned
Richardson iteration routine,
whose pseudocode is given in Figure~\ref{fig:randRichardson}.

\begin{figure}
\vskip 0.2in
\centering
\fbox{
\begin{minipage}{6in}
\noindent $\vecx = \textsc{RandRichardson} (G, T, \er,\textsc{Solver}, \vecb, \epsilon)$,
where $G$ is a graph, $T$ is a tree, $\er$ are upper bounds of
the stretches of edges of $G$ w.r.t. $T$,
$\vecb$ is the vector to be solved, and $\epsilon$ is the target error.
\begin{enumerate}
\item Set $\epsilon_1 = \frac{1}{320 c_s \log{n}}$ and $t = \Oh\left(\log\left( \epsilon^{-1} \log{n}\right)\right)$.
\item Let $\matz$ be the linear operator corresponding to the solver given in Lemma~\ref{lem:coarseSolver}
\item Repeat
\begin{enumerate}
\item $\vecx_0 = 0$.
\item For $i = 1 \ldots t$
\begin{enumerate}
\item $(H_i, T_i, \er_i) = \textsc{RandPrecon}(G, T, \er, \delta)$.
\item $\vecr_i = \laplacian_G \vecx_{i - 1} - \vecb$.
\item $\vecy_i = \textsc{Eliminate\&Solve}\left(H_i, T_i, \er_i,\textsc{Solver},  \vecr_i, \epsilon_1 \right)$.
\item $\vecx_{i} = \vecx_{i - 1} - \frac{1}{10} \vecy_i$.
\end{enumerate}
\end{enumerate}
\item Until $\nbr{\matz\left( \laplacian_G \vecx_t - \vecb \right)}_{\laplacian_G} \leq \frac{\epsilon}{c_Z \log^4 n} \nbr{\matz \vecb}_{\laplacian_G}$.
\item Return $\vecx_t$
\end{enumerate}
\end{minipage}
}
\caption{Randomized Richardson Iteration}
\label{fig:randRichardson}
\end{figure}

The guarantees of this routine is as follows.

\begin{lemma}
\label{lem:randRichardson}
Given a Laplacian solver $\textsc{Solver}$, any graph-tree pair $(G, T)$, bounds on stretch $\er$,
vector $\vecb = \laplacian_G \bar{\vecx}$ and error $\epsilon > 0$,
$\textsc{RandRichardson}(G, T, \er, \textsc{Solver}, \vecb, \epsilon)$ returns
with high probability a vector $\vecx$ such that
\begin{align}
	\nbr{\vecx - \bar{\vecx}}_{\laplacian_G}
	& \leq \epsilon \nbr{\bar{\vecx}}_{\laplacian_G},
\label{eq:richardsonGood}
\end{align}
and the algorithm takes an expected $\Oh(\log(\epsilon^{-1}) + \log\log{n}))$
iterations. Each iteration consists of one call to $\textsc{Solver}$
on a graph with $\Oh( \nbr{\er}_{p}^{p})$ edges and error $\frac{1}{\Oh(\log{n})}$,
plus an overhead of $\Oh(m + \nbr{\er}_{p}^{p})$ operations.
\end{lemma}

\Proof
Consider each iteration step using the preconditioner $H_i$ generated by \textsc{RandPrecon}.
The error reduction given in Lemma~\ref{lem:randPrecon} gives:
\begin{align*}
\expct{H}{\nbr{\bar{\vecx} - \left( \vecx_{i - 1} - \frac{1}{10} \laplacian_H^{\dag} \vecr_i \right)}_{\laplacian_G}}
\leq \left( 1 - \frac{1}{160} \right) \nbr{\bar{\vecx} - \vecx_{i - 1}}_{\laplacian_G}.
\end{align*}
On the other hand, the guarantee for $\textsc{Solver}$ gives
\begin{align*}
\nbr{\vecy_i - \laplacian_H^{\dag} \vecr_i}_{\laplacian_H}
& \leq \epsilon_1 \nbr{\laplacian_H^{\dag} \vecr_i}_{\laplacian_H}.
\end{align*}
Substituting in the spectral bound between $\laplacian_G$ and $\laplacian_H$
given by Lemma~\ref{lem:spectralBound} in turn gives:
\begin{align*}
\nbr{\vecy_i - \laplacian_H^{\dag} \vecr_i}_{\laplacian_G}
& \leq \sqrt{c_s \log{n}} \epsilon_1 \nbr{\laplacian_G \left( \bar{\vecx} - \vecx_{i-1}\right) }_{\laplacian_H^{\dag}}\\
& \leq c_s \log{n} \epsilon_1  \nbr{ \bar{\vecx} - \vecx_{i-1} }_{\laplacian_G}\\
& \leq \frac{1}{320} \nbr{ \bar{\vecx} - \vecx_{i-1} }_{\laplacian_G}.
\end{align*}
Combining this with the above bound via the triangle inequality then gives
\begin{align*}
\expct{H}{\nbr{\bar{\vecx} - \left( \vecx_{i - 1} - \frac{1}{10} \laplacian_H^{\dag} \vecr_i \right)}_{\laplacian_G}}
& \leq \left( 1 - \frac{1}{160} \right) \nbr{\bar{\vecx} - \vecx_{i - 1}}_{\laplacian_G}
+ \frac{1}{320} \nbr{\bar{\vecx} - \vecx_{i - 1}}_{\laplacian_G}\\
& \leq \left( 1 - \frac{1}{320} \right)  \nbr{\bar{\vecx} - \vecx_{i - 1}}_{\laplacian_G}.
\end{align*}
Hence the expected error $\nbr{\bar{\vecx} - \vecx_i}$ decreases by
a constant factor per iteration.
After  $\Oh(\log(\epsilon^{-1} \log{n}))$ iterations the expected error is less than
$\frac{1}{2} \frac{\epsilon}{c_{Z} \log^{4}n}$, where $c_Z$ is the constant from Lemma~\ref{lem:coarseSolver}.
Markov's inequality gives that
\begin{align}
 \nbr{\vecx_t -  \bar{\vecx}}_{\laplacian_G} \leq\frac{\epsilon}{c_{Z} \log^4 n} \nbr{\bar{\vecx}}_{\laplacian_G}
 \label{eq:solnActuallyGood}
\end{align}
with probability at least $\frac{1}{2}$.
By lemma~\ref{lem:coarseSolver} we have w.h.p
\[
\laplacian_G^{\dag} \preceq \matz \preceq c_{Z} \log^{4}{n} \laplacian_G^{\dag}.
\]
If this equation holds, then the termination criterion is satisfied whenever
equation~\ref{eq:solnActuallyGood} holds, because
\begin{align*}
\nbr{\matz \left( \laplacian_G \vecx_t - \vecb \right) }_{\laplacian_G}
&\leq c_{z} \log^{4}{n} \nbr{\vecx_t -  \bar{\vecx}}_{\laplacian_G} \\
& \leq \epsilon \nbr{\bar{\vecx}}_{\laplacian_G}\\
& \leq \epsilon \nbr{\matz \vecb}_{\laplacian_G}.
\end{align*}
On the other hand, when the termination criterion holds,
\begin{align*}
\nbr{ \bar{\vecx} - \vecx_t  }_{\laplacian_G}
& \leq  \nbr{ \laplacian_G \left( \bar{\vecx} - \vecx_t \right) }_{\matz} \\
& \leq \nbr{\matz \left( \laplacian_G \vecx_t - \vecb \right) }_{\laplacian_G}\\
& \leq \frac{\epsilon}{c_Z \log^4 n} \nbr{\matz \vecb}_{\laplacian_G} \\
& \leq  \epsilon \nbr{ \bar{\vecx} }_{\laplacian_G}.
\end{align*}
This means that w.h.p. equation~\ref{eq:richardsonGood} is satisfied
when the algorithm terminates, and the algorithm terminates with
probability at least $\frac{1}{2}$ on each iteration.
So the expected number of iterations of the outer loop is $\Oh(1)$.
\QED

It remains to give use this routine recursively.
We correct for the errors of introducing scaling factors into the tree
using preconditioned Chebyshev iteration.

\begin{restatable}[Preconditioned Chebyshev Iteration]{lemma}{preconCheby}
\label{lem:preconCheby}
Given a matrix $\mata$ and a matrix $\matb$ such that $\mata \preceq \matb \preceq \kappa \mata$
for some constant $\kappa > 0$,
along with error $\epsilon > 0$ and
a routine $\textsc{Solve}_\matb$ such that for any vector $\vecb$ we have
\begin{align*}
\nbr{\textsc{Solve}_{B}(\vecb) - B^{\dag} \vecb}_{\matb} \leq \frac{\epsilon^{4}}{30 \kappa^{4}} \nbr{\vecb}_{B^{\dag}};
\end{align*}
preconditioned Chebyshev iteration gives a routine
$\textsc{Solve}_{\mata}(\cdot)=\textsc{PreconCheby} \left(\mata,\matb,\textsc{Solve}_{\matb},\cdot\right),$ such that in the exact arithmetic model,
for any vector $\vecb$,
\begin{itemize}
\item 
\begin{align*}
\nbr{\textsc{Solve}_{\mata}(\vecb) - \mata^{\dag} \vecb}_{\mata} \leq \epsilon \nbr{\vecb}_{\mata^{\dag}},
\end{align*}
 and
\item
$\textsc{Solve}_\mata(\vecb)$ takes $O(\sqrt{\kappa} \log( 1 / \epsilon) )$ iterations, each consisting
of one call to $\textsc{Solve}_\matb$ and a matrix-vector multiplication using $\mata$.
\end{itemize}
\end{restatable}

\begin{figure}
\vskip 0.2in
\centering
\fbox{
\begin{minipage}{6in}
\noindent $\vecx = \textsc{Solve} (G, T, \er, \vecb, \epsilon)$,
where $G$ is a graph, $T$ is a tree, $\er$ are upper bounds of
the stretches of edges in $G$ w.r.t. $T$,
$\vecb$ is the vector to be solved, and $\epsilon$ is the goal error.
\begin{enumerate}
\item Set $\kappa = c (\log \log n)^{4/(2p-1)} \left( \frac{\nbr{\er}_p^{p}}{m} \right)^{1/p}$ for an appropriate constant $c$ (dependent on $p$).
\item Let $(H, T', \er')$ be the graph-tree tuple with $T$ scaled up by a factor of $\kappa$, $\er$ scaled down by a factor of $\kappa$.
\item $\vecx = \textsc{PreconCheby} \left(G, H,
\textsc{RandRichardson}(H, T', \er',\textsc{Solve}, \epsilon^{4} \kappa^{-4}), \vecb \right)$.
\item Return $\vecx$
\end{enumerate}
\end{minipage}
}
\caption{Recursive Solver}
\label{fig:solver}
\end{figure}

The pseudocode of our algorithm is given in Figure~\ref{fig:solver}.
Below we prove its guarantee.

\begin{lemma}
\label{lem:solver}
Given a parameter $1/2 < p < 1$ and a graph-tree tuple $(G, T, \er)$
with $m$ edges such that $\nbr{\er}_p^{p} \leq m \log^{p}{n}$.
For any vector $\vecb = \laplacian_G \bar{\vecx}$,
$\textsc{Solve}(G, T, \er, \vecb, \frac{1}{320 c_s \log{n}})$ returns w.h.p. a vector $\vecx$ such that
\begin{align*}
\nbr{\bar{\vecx} - \vecx}_{\laplacian_G} \leq \frac{1}{320 c_s \log{n}} \nbr{\vecx}_{\laplacian_G},
\end{align*}
and its expected running time is
\begin{align*}
\Oh\left(m \left( \frac{\nbr{\er}_p^{p}}{m} \right)^{\frac{1}{2p}} \log\log^{2 + \frac{2}{2p-1}} {n} \right).
\end{align*}
\end{lemma}

\Proof
The proof is by induction on graph size. As our induction hypothesis, we assume the lemma to be true for all graphs of size $m' < m$. 
The choice of $\kappa$ gives
\begin{align*}
\nbr{\er'}_p^{p} \leq \frac{m}{c^p \log\log^{2 + \frac{2}{2p-1}} {n}}.
\end{align*}
The guarantees of randomized Richardson iteration from
Lemma~\ref{lem:randRichardson} gives that all the randomized preconditioners have both off-tree edge count and off-tree stretch
bounded by $\Oh(\nbr{\er'}_p^{p}) = \Oh \left( \frac{m}{c^p \log\log^{2 + \frac{2}{2p-1}} {n}}  \right) $.

An appropriate choice of $c$ makes both of these values strictly less
than $m$, and this allows us to apply the inductive hypothesis on the graphs obtained
from the randomized preconditioners by $\textsc{Eliminate\&Solve}$.

As $\kappa$ is bounded by $c \log^2{n}$ and $\epsilon$ is set to
$\frac{1}{320 c_s \log{n}}$, the expected cost of the recursive calls made by
\textsc{RandRichardson} is
\begin{align*}
\Oh(m \log\log {n}).
\end{align*}
Combining this with the iteration count in $\textsc{PreconCheby}$ of
\[
\Oh(\sqrt{\kappa} \log( 1 / \epsilon)) = \Oh \left( (\log \log n)^\frac{2}{2p-1} \left( \frac{\nbr{\er}_p^{p}}{m} \right)^{\frac{1}{2p}} \log \log n \right)
\]
gives the inductive hypothesis.
\QED
\\ \\

To prove theorem~\ref{thm:mainSDD}, we first invoke $\textsc{Solve}$ with
$\epsilon$ set to a constant.
Following an analysis identical to the proof of lemma~\ref{lem:solver}, at the
top level each iteration of $\textsc{PreconCheby}$ will require $\Oh( m \log\log {n} )$ time, but now only
\[
\Oh(\sqrt{\kappa} \log( 1 / \epsilon)) = \Oh \left( (\log \log n)^\frac{2}{2p-1} \left( \frac{\nbr{\er}_p^{p}}{m} \right)^{\frac{1}{2p}}\right)
\]
iterations are necessary.
Setting $p$ arbitrarily close to $1$ means that for any constant $\delta > 0$
and relative error $\epsilon$, there is a solver for $\LL_{G}$ 
that runs in $\Oh(m\log^{1/2}n \log\log^{3+\delta}n)$ time.
This error can be reduced using Richardson iteration as stated below.

\begin{restatable}{lemma}{preconRichardson}
\label{lem:preconRichardson}
If $\AA$, $\BB$ are matrices such that $\AA \preceq \BB \preceq 2 \AA$
and $\textsc{Solve}_{\BB}$ is a routine such that for any vector $\vecb$,
we have $\nbr{\textsc{Solve}_{\BB}(\vecb)- \BB^{\dag} \vecb}_{\BB}
\leq \frac{1}{5} \nbr{\BB^{\dag} \vecb}_{\BB}$, then
there is a routine $\textsc{Solve}_{\AA,\epsilon}$ which runs in $\Oh(c_{\alpha}\log(\frac{1}{\epsilon}))$ iterations with the guarantee that for any vector $\vecb$ we have $\nbr{\textsc{Solve}_{\AA,\epsilon}(\vecb)- \mata^{\dag}\vecb}_{\AA} \leq \epsilon \nbr{\AA^{\dag} \vecb}_{\AA}$.
Each iteration involves one call to $\textsc{Solve}_{\BB}$,
a matrix-vector multiplication involving $\AA$ and $O(1)$ arithmetic operations on vectors.
\end{restatable}

We will use Richardson iteration as the outer loop, while transferring
solutions and errors to the original graph using the guarantees of the
embeddable tree given in Lemma~\ref{lem:hasTree}.

\Proofof{Theorem~\ref{thm:mainSDD}}
Using Fact~\ref{fact:errorTransfer} on the solver described above
for $\LL_G$ gives a solver for
$(\PPi_1 \PPi \LL_{G}^{\dag}  \PPi^T \PPi_1^T)^{\dag}$ with
relative error $\frac{1}{5}$.
This condition and Lemma~\ref{lem:hasTree} Part~\ref{item:goodApprox}
then allows us to invoke the above Lemma with $\AA = \LL_{\hat{G}}$
and $\BB = \PPi_1 \PPi \LL_{G}^{\dag}  \PPi^T \PPi_1^T$.
Incorporating the $O(\log(\frac{1}{\epsilon}))$ iteration count
and the reduction from SDD linear systems then gives the overall result.
\QED

\section*{Acknowledgements}

We thank Jon Kelner, Gary Miller, and Dan Spielman for their advice, comments and discussions.

\bibliographystyle{alpha}
\bibliography{references}

\begin{appendix}

\section{Chebyshev Iteration with Errors}
\label{sec:cheby}

We now check that preconditioned Chebyshev iteration can tolerate
a reasonable amount of error in each of the calls to the preconditioner.
A more detailed treatment of iterative methods can be found in
the book by Trefethen and Bau~\cite{TrefethenB97:book}.
Our presentation in this section is geared to proving the following guarantee.

\preconCheby

As the name suggests, Chebyshev iteration is closely related
with Chebyshev Polynomials.
There are two kinds of Chebyshev Polynomials, both defined
by recurrences.
Chebyshev polynomials of the first kind, $T_n(x)$ can be defined as:
\begin{align*}
T_0(x)
&= 1,\\
T_1(x)
&= x,\\
T_{i + 1}(x)
&= 2x T_{i}(x) - T_{i - 1}(x).
\end{align*}

Preconditioned Chebyshev iteration is given by the following
recurrence with $\delta$ set to $1 + \frac{1}{\kappa}$:

\begin{figure}
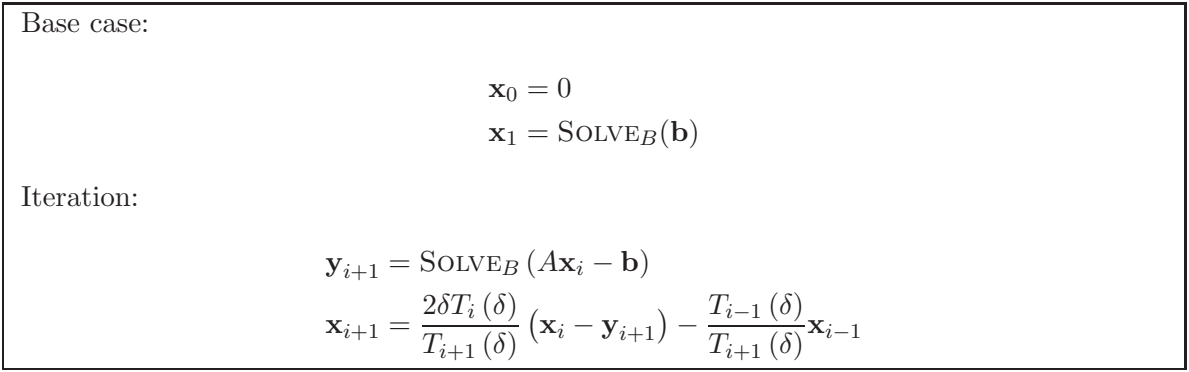

\vskip 0.2in
\noindent
\fbox{
\begin{minipage}{6in}

Base case:
\begin{align*}
\vecx_0
&= 0\\
\vecx_1
&= 	\textsc{Solve}_\matb(\vecb)
\end{align*}

Iteration:
\begin{align*}
\vecy_{i + 1}
& = \textsc{Solve}_\matb \left( \mata \vecx_i - \vecb \right)\\
\vecx_{i + 1}
& = \frac{2\delta T_{i}\left( \delta \right)}
		{T_{i+1}\left( \delta \right)} \left( \vecx_{i} - \vecy_{i + 1} \right)
	- \frac{T_{i - 1}\left( \delta \right)}
		{T_{i+1}\left(\delta \right)} \vecx_{i - 1}
\end{align*}

\end{minipage}
}
\vskip 0.2in
\caption{Preconditioned Chebyshev Iteration}
\label{fig:preconCheby}
\end{figure}

To bound the convergence of this iteration, it is helpful to
use the following closed form for $T_i(x)$:
\begin{align*}
T_i(x)
& = \frac{\left( x - \sqrt{x^2 - 1}  \right)^i
	+ \left( x + \sqrt{x^2 - 1}  \right)^i}{2}.
\end{align*}
The following facts about Chebyshev polynomials of the
first kind will be used to bound convergence.
\begin{fact}
\label{fact:cheby1Cos}
If $x = \cos(\theta)$, then
\begin{align*}
T_i(x)
& = \cos(i \theta).
\end{align*}
\end{fact}
This implies that if $|x| \leq 1$, $|T_i(x)| \leq 1$, and we will
pass the error of the algorithm through it.
For convergence, we also need the opposite statement
for lower bounding $T_n(x)$ when $x$ is large.
\begin{fact}
\label{fact:cheby1Lower}
If $x = 1 + \frac{1}{\kappa}$, then:
\begin{align*}
T_i(x)
& \geq \frac{1}{2} \left( x + \sqrt{x^2 - 1}  \right)^i,\\
& \geq \frac{1}{2} \left( 1 + \sqrt{1 + \frac{2}{\kappa} - 1 }  \right)^i,\\
& \geq \frac{1}{2} \left( 1 + \frac{1}{\sqrt{\kappa}}  \right)^i.
\end{align*}
\end{fact}

We can also show that these terms are steadily increasing:
\begin{fact}
\label{fact:cheby1Ratio}
If $i \leq j$ and $x \geq 1$, then $T_i(x) \geq \frac{1}{2} T_j(x)$.
\end{fact}

\Proof
$x \geq 1$ implies $0 \leq  x - \sqrt{x^2 - 1}  \leq 1$
and $ 1 \leq x + \sqrt{x^2 - 1}$.
Therefore
\begin{align*}
T_{i + 1} \left( x \right)
& \geq \frac{\left( x + \sqrt{x^2 - 1}  \right)^{i + 1}}{2},\\
& \geq \frac{\left( x + \sqrt{x^2 - 1}  \right)^{i}}{2},\\
& \geq T_{i} \left( x \right) - \frac{1}{2}.
\end{align*} 
Fact~\ref{fact:cheby1Lower} also gives $T_{i + 1} (x) \geq \frac{1}{2}$.
Combining these gives $T_{i} (\delta) \geq \frac{1}{2} T_j( \delta )$.
\QED

The errors given by $\textsc{Solve}_{\matb}$ will accumulate over the iterations.
To bound them, we need Chebyshev polynomials of the second kind.
These polynomials, $U_n(x)$, follow the same recurrence
but have a different base case:
\begin{align*}
U_{-1}(x)
&= 0,\\
U_0(x)
&= 1,\\
U_{i + 1}(x)
&= 2x T_{i}(x) - T_{i - 1}(x).
\end{align*}
Chebyshev polynomials of the second kind are related to
Chebyshev polynomials of the first kind by the following identity:
\begin{fact}
\label{fact:relationfirstsecond}
\begin{align*}
U_i(x)
&= 
\begin{cases}
2 \sum_{j \leq i \text{ odd}} T_{j}(x)& \text{If $i$ is odd}, and\\
\left ( 2 \sum_{j \leq i \text{ even}} T_{j}(x) \right ) - 1 & \text{If $i$ is even}.\\
\end{cases}
\end{align*}
\end{fact}
Since $T_0 = 1$, and $|T_j(x)| \leq 1$ whenever $x \leq 1$, this implies
\begin{fact}
\label{fact:secondbound}
For all $x$ satisfying $|x| \leq 1$,
\begin{equation*}
|U_i(x)| \leq i+1
\end{equation*}
\end{fact}

We will let the deviation caused by $\textsc{Solve}_\matb$ at iteration
$i$ to be $\err_i$, giving
\begin{align*}
\vecy_{i + 1} & = \matb^{\dag} \left( \mata \vecx_i - \vecb \right) + \err_i
\end{align*}
where $\nbr{\err_i}_{\matb} \leq \nbr{ \mata \vecx_i - \vecb }_{\matb^{\dag}}$.
To analyze the recurrence, it is crucial to consider the matrix
\begin{align*}
X & = \delta \left( \mati -
\mata^{1/2} \matb^{\dag} \mata^{1/2} \right).
\end{align*}
The given condition of $\mata \preceq \matb \preceq \kappa \mata$ gives
\begin{align*}
\frac{1}{\kappa} \mati
\preceq \mata^{1/2} \matb^{\dag} \mata^{1/2} \preceq \mati,
\end{align*}
which when combined with the setting of $\delta = 1 + \frac{1}{\kappa}$
gives
\begin{align*}
0 \preceq X \preceq I.
\end{align*}
Fact~\ref{fact:cheby1Cos} then gives that $T_i(X)$ has all
eigenvalues between $[-1, 1]$.
This `shrinkage' property is key to our analysis.

We can show that the deviation between $\vecx_i$ and $\bar{\vecx} = \mata^{\dag}b$
behaves according to Chebyshev polynomials of the first kind in $X$
while the errors accumulate according to
Chebyshev polynomials of the second kind in $X$.

\begin{lemma}
\label{lem:totalerror}
If $\bar{\matx} = \mata^{\dag} b$, then at iteration $i$ we have
\begin{align*}
T_{i}\left( {\delta} \right)
\left( \vecx_i - \bar{\vecx} \right)
&= \mata^{\dag 1/2}  T_{i} \left( \matx \right)  \mata^{1/2} \bar{\vecx}
+ 2 \delta \sum_{j = 1}^{i}  T_{j - 1} \left( \delta \right) 
   \mata^{\dag 1/2}  U_{i - j} \left( \matx \right) \mata^{1/2} \err_j,
\end{align*}
where $X = \delta \left( \mati - \mata^{1/2} \matb^{\dag} \mata^{1/2} \right)$ and
$T_i(X)$ and $U_i(x)$ are Chebyshev polynomials of the first and second kind respectively
\end{lemma}

\Proof
The proof is by induction.

The base case can be checked as follows:
\begin{align*}
\bar{\vecx} - \vecx_0
&= \bar{\vecx},\\
&= \mata^{\dag 1/2} \mata^{1/2} \bar{\vecx};\\
\bar{\vecx} - \vecx_1
&= \bar{\vecx} - \matb^{\dag} \vecb + \err_1,\\
& = \bar{\vecx} - \matb^{\dag} \mata \bar{\vecx},\\
&= \mata^{\dag 1/2} \left( \mati - \mata^{1/2} \matb^{\dag} \mata^{1/2} \right) \mata^{1/2} \bar{\vecx}
+ \mata^{\dag 1/2} \mata^{1/2} \err_1.
\end{align*}

For the inductive case,
the recurrence can be rearranged to give:
\begin{align*}
T_{i+1}\left( \delta \right) \vecx_{i + 1}
& = 2 \delta T_{i} \left( \delta \right)
	\left( \vecx_{i} - \vecy_{i + 1} \right)
	- T_{i - 1} \left( \delta \right)
		\left( \delta \right) \vecx_{i - 1}
\end{align*}

Recall from the definition of Chebyshev polynomials
of the first kind that:
\begin{align*}
T_{i+1}\left( \delta \right)
&= 2 \left( \delta \right) T_{i} \left( \delta \right) 
	- T_{i - 1} \left( \delta \right)
\end{align*}
So we can subtract both sides from
$T_{i+1}\left( \delta \right) \bar{\vecx}$ to get: 
\begin{align*}
T_{i+1}\left({\delta} \right) \left(\bar{\vecx} - \vecx_{i + 1}  \right)
& = 2 \delta T_{i} \left({\delta} \right) \left( \bar{\vecx}_i - \vecx_i + \vecy_i \right)
	- T_{i - 1} \left( {\delta} \right) \left( \bar{\vecx}  -  \vecx_{i - 1}\right)
\end{align*}
The change, $y_i$, can be viewed as computed by multiplying
the difference at iteration $i$ by $\matb^{-1} \mata$,
plus the error vector $\err_i$:
\begin{align*}
y_{i + 1}
& = \matb^{\dag} \left( \mata \vecx_i - \vecb \right) + \err_{i + 1} \\
& = \matb^{\dag} \left( \mata \vecx_i - \mata \bar{\vecx} \right) + \err_{i + 1} \\
& = \matb^{\dag} \mata \left( \vecx_i - \bar{\vecx} \right)  + \err_{i + 1}
\end{align*}
Combining this gives
\begin{align*}
T_{i+1}\left({\delta} \right) \left(\bar{\vecx} - \vecx_{i + 1}  \right)
& = 2 \delta T_{i} \left({\delta} \right) \left( \mati - \matb^{\dag} \mata \right) \left( \bar{\vecx} - \vecx_i\right)
	- T_{i - 1} \left( {\delta} \right) \left( \bar{\vecx}  -  \vecx_{i - 1}\right)
	 + 2 \delta T_{i}\left( \delta \right) \err_{i + 1}\\
& = 2 \mata^{\dag 1/2} \matx \mata^{1/2} T_{i} \left({\delta} \right) \left( \bar{\vecx} - \vecx_i\right)
	- T_{i - 1} \left( {\delta} \right) \left( \bar{\vecx}  -  \vecx_{i - 1}\right)
	 + 2 \delta T_{i} \left(\delta \right) \err_{i + 1}.
\end{align*}

From this, we can then show the inductive case by collecting all the terms
and checking that the coefficients satisfy
the recurrences for Chebyshev polynomials.
Substituting in the inductive hypothesis gives:
\begin{align*}
T_{i + 1}\left({\delta} \right) \left( \bar{\vecx} - \vecx_i \right)
& = 2 \mata^{\dag1/2} \matx \mata^{1/2} \left( \mata^{\dag 1/2} T_{i} \left( \matx \right)  \mata^{1/2} \bar{\vecx}
+ 2 \delta \sum_{j = 1}^{i}  T_{j - 1} \left( \delta \right) 
   \mata^{\dag 1/2}  U_{i - j} \left( \matx \right) \mata^{1/2} \err_j\right)
\\ & \qquad
+ \mata^{\dag 1/2}  T_{i - 1} \left( \matx \right)  \mata^{1/2} \bar{\vecx}
+ 2 \delta \sum_{j = 1}^{i - 1}  T_{j - 1} \left( \delta \right) 
   \mata^{\dag 1/2}  U_{i - 1 - j} \left( \matx \right) \mata^{1/2} \err_j
	+ 2 \delta T_{i}\left( \delta \right) \err_{i + 1}
\end{align*}
Since $\mata$, $\matb$ and $\matx$ share the same null space and the first
term is left-multiplied by $\mata^{\dag1/2}$,
the $\mata^{1/2}$ and $\mata^{\dag 1/2}$ terms cancel with each other.
Collecting the terms according to $\bar{\vecx}$ and $\err_j$ then gives
\begin{align*}
T_{i + 1}\left({\delta} \right) \left( \bar{\vecx} - \vecx_i \right)
& = \mata^{\dag 1/2} \left( 2 \matx T_i\left(\matx\right) - T_{i - 1} \left( \matx \right) \right) \mata^{1/2} \bar{\vecx}
\\ & \qquad
+ 2 \delta \sum_{j = 1}^{i }  T_{j - 1}\left(\delta\right)
	\mata^{\dag 1/2} \left( 2 \matx U_{i - j} \left( \matx \right) - U_{i - 1 - j} \left( \matx \right) \right) \mata^{1/2} \err_j
+ 2 \delta T_{i}\left( \delta \right)  \err_{i + 1}\\
& = \mata^{\dag 1/2} T_{i + 1}\left(\matx\right) \mata^{1/2} \bar{\vecx}
+ 2 \delta \sum_{j = 1}^{i }  T_{j - 1}\left(\delta\right)
  U_{i + 1 - j} \left( \matx \right) \mata^{1/2} \err_j + 2 \delta T_{i}\left( \delta \right) \err_{i + 1}
\end{align*}

As $U_{-1}(x) = 0$, we can also include in the $j = i$ term
in the summation of error terms.
So the inductive hypothesis holds for $i + 1$ as well.
\QED

The bound on Chebyshev polynomials of the second kind (Fact~\ref{fact:secondbound}) then allows us to bound
the error in the $\mata$-norm.

\begin{lemma}
\label{lem:totalDeviation}
The accumulation of errors after $i$ iterations can
be bounded by:
\begin{align*}
\nbr{\bar{\vecx} - \vecx_i}_{\mata}
& \leq \frac{1}{T_i \left( \delta\right) } \nbr{\bar{\vecx}}_{\mata} +
\sum_{j = 1}^{i} 6i \nbr{\err_j}_{\mata}
\end{align*}
\end{lemma}

\Proof
By the identity proven in Lemma~\ref{lem:totalerror} above,
and the property of norms, we have:
\begin{align*}
\nbr{\bar{\vecx} - \vecx_i}_{\mata}
& = \nbr{\mata^{\dag 1/2} T_{i + 1}\left(\matx\right) \mata^{1/2} \bar{\vecx}
+ 2 \delta \sum_{j = 1}^{i - 1}  T_{j - 1}\left(\delta\right)
  U_{i + 1 - j} \left( \matx \right) \mata^{1/2} \err_j + 2 \delta T_{i}\left( \delta \right) \err_{i + 1}}_{\mata}\\
& = \frac{1}{T_{i} \left( {\delta} \right)}
	\nbr{T_{i + 1}\left( \matx \right) \mata^{1/2} \bar{\vecx}
+ 2 \delta \sum_{j = 1}^{i}  T_{j - 1}\left(\delta\right)
  U_{i + 1 - j} \left( \matx \right) \mata^{1/2} \err_j }_{2},
\end{align*}
on which triangle inequality gives:
\begin{align*}
\nbr{\bar{\vecx} - \vecx_i}_{\mata}
& \leq  \frac{1}{T_{i} \left( {\delta} \right)}
	\nbr{T_{i + 1}\left(\matx\right) \mata^{1/2} \bar{\vecx}}_2
+ \frac{2 \delta  T_{j - 1}\left(\delta\right)}{T_{i} \left( {\delta} \right)}
\sum_{j = 1}^{i} \nbr{ U_{i - j} \left( \matx \right) \mata^{1/2} \err_j }_{2}
\end{align*}

The upper bound on $T$ implies that the eigenvalues of $T_i(\matx)$ all have absolute value at most 1; similarly the upper bound on $U$ given in Fact~\ref{fact:secondbound} implies that all eigenvalues of $U_{k}(\matx)$ have absolute value at most $k+1$.  This implies that for any vector $\vecx$, $\nbr{ T_i(\matx) \vecx }_2 \leq \nbr{ \vecx }_2$ and $\nbr{ U_k(\matx) \vecx }_2 \leq (k+1) \nbr{ \vecx }_2$.  Furthermore, by Fact~\ref{fact:cheby1Lower}, $\frac{2 \delta T_{j - 1}\left(\delta\right)} < 6$.  Applying these bounds, and the definition of $\mata$-norm, gives
\begin{align*}
\nbr{\bar{\vecx} - \vecx_i}_{\mata}
& \leq  \frac{1}{T_{i} \left( {\delta} \right)} \nbr{\bar{\vecx}}_\mata + 6 \sum_{j = 1}^{i} (i-j+1) \nbr{ \err_j }_{\mata} \\
& \leq \leq  \frac{1}{T_{i} \left( {\delta} \right)} \nbr{\bar{\vecx}}_\mata + 6 \sum_{j = 1}^{i} i \nbr{ \err_j }_{\mata}
\end{align*}
\QED

As the error bound guarantee of $\textsc{Solver}_\matb$ is relative,
we need to inductively show that the total error is small.
This then leads to the final error bound.

\Proofof{Lemma~\ref{lem:preconCheby}}
The proof is by induction.
We show that as long as $i < \kappa \epsilon^{-1}$, we have
\begin{align*}
\nbr{\bar{\vecx} - \vecx_i}_{\mata}
\leq & \left( \frac{1}{T_{i} \left( {\delta} \right)}
+   \frac{\epsilon^2 i}{2 \kappa} \right) \nbr{\bar{\vecx}}_{\mata},
\end{align*}
and $\nbr{\err_{j}}_{\mata} \leq
\frac{\epsilon^3}{24 \kappa^2} \nbr{\bar{\vecx}}_{\mata}$
for all $j \leq i$.

The base case of $i = 0$ follows from $T_{0}( {\delta} ) = 0$.
For the inductive case, suppose the result is true for $i - 1$.
Then as $i < \kappa \epsilon^{-1}$ and $T_i (\delta) \geq 1$, we have
$\nbr{\bar{\vecx} - \vecx_{i - 1}}_{\mata} \leq 2 \nbr{\bar{\vecx}}_{\mata}$.
As the vector passed to $\textsc{solve}_{\matb}$ is $\mata \vecx_{i - 1} - \vecb
= \mata(\vecx_{i - 1} - \bar{\vecx})$ and $\matb^{\dag} \preceq \mata^{\dag}$, we have
\begin{align}
\nbr{\mata\left(\vecx_{i - 1} - \bar{\vecx}\right)}_{\matb^{\dag}}
& = \sqrt{\left(\vecx_{i - 1} - \bar{\vecx}\right)^T \mata \matb^{\dag} \mata \left(\vecx_{i - 1} - \bar{\vecx}\right)}\\
& \leq \sqrt{\left(\vecx_{i - 1} - \bar{\vecx}\right)^T \mata \left(\vecx_{i - 1} - \bar{\vecx}\right)}\\
& = \nbr{\vecx_{i - 1} - \bar{\vecx}}_{\mata}\\
& \leq 2 \nbr{\bar{\vecx}}_{\mata}.
\end{align}
Therefore the guarantees of $\textsc{Solver}_{\matb}$ gives
$\nbr{\err_{i}}_{\matb} \leq \frac{\epsilon^{3} }{48 \kappa^2}$.
Combining this with $\mata \preceq \matb$ gives the bound on $\err_i$.

Substituting these bounds into Lemma~\ref{lem:totalDeviation}
in turn gives the inductive hypothesis for $i$.
The lower bound on $T_i(\delta)$ gives that when
$i = \Oh( \sqrt{\kappa} \log( 1 / \epsilon ) )$,
the first term is less than $\frac{\epsilon}{2}$.
As $\log( 1 / \epsilon ) \leq \frac{1}{\epsilon}$,
the second term can be bounded by $\frac{\epsilon}{2}$ as well.
Combining these two error terms gives the overall error.
\QED

We remark that the exponent on $\kappa$ and $\epsilon$ in this analysis
are not tight, and will be improved in a future version.

\section{Finding Electrical Flows}
\label{sec:electrical}

We now show that the solver given in Theorem~\ref{thm:mainSDD}
can also be used to find electrical flows in similar time.
This problem can be viewed as the dual of computing vertex potentials,
and is the core problem solved in the flow energy reduction based algorithms
by Kelner et al.~\cite{KelnerOSZ13} and Lee and Sidford~\cite{LeeS13}.
As flows to defined on the edges of graphs instead of vertices,
it is helpful to define the edge vertex incidence matrix.

\begin{definition}
\label{def:edgeVertex}
The edge-vertex incidence matrix of a weighted graph $G = (V, E)$ is given by
\[
\BB_{e, u} = 
\begin{cases}
1 &\text{if $u$ is the head of $e$}\\
-1 &\text{if $u$ is the tail of $e$}\\
0 &\text{otherwise}
\end{cases}
\]
\end{definition}
It can be checked that if $\RR$ is the diagonal matrix containing all the
resistances, the graph Laplacian is given by $\LL = \BB^T \RR \BB$.

Given a flow $\ff$, its residual at vertices is given by $\BB^T \ff$.
Also, the energy of the flow is given by $\eFlow{\ff} = \norm{\ff}_{\RR}$.
The electrical flow problem is finding the minimum energy flow whose
residue meets a set of demands $\dd$.
It can be characterized as follows.

\begin{fact}
\label{fact:er}
For a demand $\dd$, the minimum energy electrical flow $\bar{\ff}$
is given by
\[
\bar{\ff} = \RR^{-1} \BB \LL^{\dag} \dd,
\]
and its energy, $\eFlow{\bar{\ff}}$ equals to $\norm{\dd}_{\LL^{\dag}}$.
\end{fact}

As a result, a natural algorithm for computing a flow that approximately
minimizes electrical energy is to solve for approximate potentials $\LL^{\dag} \dd$.
Previous reductions between these problems such as the one by
Christiano et al.~\cite{ChristianoKMST11} ran the solver to high accuracy
to recover these potentials.
Then any difference between the residue and demands are fixed combinatorially.
Here we show that this exchange can happen with low error in a gradual fashion.
The following lemma is the key to our algorithm.

\begin{lemma}
\label{lem:energyClose}
If $\xx$ is a vector such $\norm{\xx - \LL^{\dag} \dd}_{\LL}
\leq \epsilon \norm{\dd}_{\LL^{\dag}}$, then
$\ff = \RR^{-1} \BB \xx$ is a flow such that
$\eFlow{\ff} \leq (1 + \epsilon) \norm{\dd}_{\LL^{\dag}}$, and
 the energy required to send the flow
$\dd - \BB \ff$ is at most $\epsilon \norm{\dd}_{\LL^{\dag}}$.
\end{lemma}

\Proof

Both steps can be checked algebraically.
For the energy of the flow, we have
\[
\eFlow{\ff}^2
=  (\RR^{-1} \BB \xx)^T \RR  (\RR^{-1} \BB \xx)
= \xx^T \LL \xx
= \norm{\xx}_{\LL}^2.
\]
Combining this with the error guarantees gives
\[
\eFlow{\ff}
= \norm{\xx}_{\LL}
\leq \norm{\LL^{\dag} \dd}_{\LL} + \norm{\xx - \LL^{\dag} \dd}_{\LL}
\leq (1 + \epsilon) \norm{\dd}_{\LL^{\dag}}.
\]

For the energy needed to reroute the demands,
note that $\BB \ff = \LL \xx$.
Substituting this in gives:
\[
\norm{\BB \ff -\dd}_{\LL^{\dag}}
= \norm{ \LL \xx - \dd}_{\LL^{\dag}}
= \norm{\xx - \LL^{\dag} \dd}_{\LL}
= \epsilon \norm{\dd}_{\LL^{\dag}}.
\]
\QED

This means that we can solve the resulting re-routing
problem to a much lower accuracy.
This decrease in accuracy in turn allows us to change our graph,
leading to a faster running time for this correction step.
We give an outline this procedure below, and will have a
more detailed exposition in the full version.

\begin{claim}
\label{claim:fastElectrical}
Given a graph $G = (V, E, \rr)$, a set of demands $\dd$,
and any error parameter $\epsilon > 0$
we can find in expected $O(m \log^{1/2}{n} \poly(\log\log {n}) \log(\epsilon^{-1}))$ time a flow $\ff$ such that with high probability $\ff$ meets the demands,
and $\eFlow{\ff} \leq (1 + \epsilon) \norm{\dd}_{\LL^{\dag}}$.
\end{claim}

\Proof
Consider running the solver given Theorem~\ref{thm:mainSDD}
to an accuracy of $\frac{\epsilon}{\log^{3}n}$, and using the
resulting flow $\ff$.
Lemma~\ref{lem:energyClose} then gives that it suffices to find another
flow with a set of demands $\dd'$ such that
$\norm{\dd'}_{\LL^{\dag}} \leq \frac{\epsilon}{\log^{3}n} \norm{\dd}_{\LL^{\dag}}$.
As the energy of $\ff$ is at most
$(1 + \frac{\epsilon}{\log^{3}n} )\norm{\dd}_{\LL^{\dag}}$,
it suffices to find a flow $\ff'$ meeting demands $\dd'$
such that $\eFlow{\ff'} \leq \frac{\epsilon}{2} \norm{\dd}_{\LL^{\dag}}$.

The fact that we can tolerate a $\frac{\log^3{n}}{2}$ factor increase
in energy in $\ff'$ allows us to find this flow on a graph with some
resistances increased by the same factor.
This allows us to reduce the value of $\nbr{\er}_{p}^{p}$ in Lemma~\ref{lem:hasTree} by a factor of about $\log^{3p}{n}$.
It can also be checked that it suffices to find electrical flows a sparsified
version of this graph.
Therefore, the solve can be ran to an accuracy of $\frac{1}{\poly(n)}$
on this smaller graph without being a bottleneck in the running time.

Adding this flow in means that we in turn need to find a flow
for some demand $\dd''$ with energy at most $\poly(n) \norm{\dd''}_{\LL^{\dag}}$.
As the relative condition number of the minimum spanning
tree with the graph can be bounded by $\poly(n)$, using it
to reroute the flow allows us to arrive at the final flow.
\QED

\section{Relation to Matrix Chernoff Bounds}
\label{sec:whp}

We now show a matrix Chernoff bounds based analysis of our
sampling routine which gives bounds that are off by log factors
on each side with high probability.
The matrix Chernoff bound that we will use is as follows:

\begin{lemma}[Matrix Chernoff, Theorem 1.1 from~\cite{Tropp12}]
\label{lem:matrixChernoff}
Let $\MM_k$ be a sequence of independent, random, self-adjoint matrices
with dimension $n$.
Assume that each random matrix satisfies
$\bvec{0} \preceq \MM_k$
and $\lambda_{\max} (\MM_k) \leq R$.
Define $\mu_{\min} = \lambda_{\min} \left( \sum_{k} \expct{}{\MM_k} \right)$
and $\mu_{\max} = \lambda_{\max} \left( \sum_{k} \expct{}{\MM_k} \right)$.
Then
\[
\prob{}{\lambda_{\min} \left( \sum_{k} \expct{}{\MM_k}\right)
\leq \left(1 - \delta\right) \mu_{\min}}
\leq n \cdot \left[ \frac{e^{-\delta}}{(1 - \delta)^{1 - \delta}}\right]^{\mu_{\min} / R}
\text{for $\delta \in [0, 1]$},
\]
and
\[
\prob{}{\lambda_{\min} \left( \sum_{k} \expct{}{\MM_k}\right)
\leq \left(1 + \delta\right) \mu_{\max}}
\leq n \cdot \left[ \frac{e^{-\delta}}{(1 - \delta)^{1 - \delta}}\right]^{\mu_{\max} / R}
\text{for $\delta \geq 0$}.
\]
\end{lemma}

As this bound is tailored for low error, we need an additional
smoothing step.
Here the fact that we add $\XX$ to the resulting sample is crucial
for our analysis.
It allows us to analyze the deviation between $\ZZ + \kappa \XX$
and $\YY + \kappa \XX$ for a parameter $\kappa$ that we will pick.
We will actually prove a generalization of both the
$\delta = \frac{1}{\Oh(\log{n})}$ case and the $\delta = \Oh(1)$ case.

\begin{lemma}
\label{lem:whp}
There exists a constant $c$ such that the
output of $\ZZ = \matz = \textsc{Sample} (\{\maty_1, \ldots, \maty_m\}, X, \er, \delta)$ satisfies with high probability
\[
\frac{1}{c \delta \log{n}} \cdot \YY
\preceq \ZZ \preceq c \delta \log{n} \cdot \YY.
\]
\end{lemma}

\Proof
Note that our sampling algorithm also picks the number of samples,
$r$, randomly between $t$ and $2 t - 1$.
However, as we can double $c$, it suffices to show the result
of taking $t$ samples is tightly concentrated.

Let $\kappa > 0$ be a parameter that we set later to about $\delta \log{n}$, and consider the
approximation between $\ZZ + \kappa \XX$ and $\YY + \kappa \XX$.
We let $\MM_1 \ldots \MM_{t}$ be the matrices corresponding to the
samples, normalized by $\YY + \kappa \XX$:
\[
\MM_{i} \defeq \frac{\delta}{\er_{i_j}} \left(\YY + \kappa \XX\right)^{-1/2} \YY_{i_j} \left(\YY + \kappa \XX\right)^{-1/2}.
\]
As all $\YY_i$s are positive semidefinite, this random matrix is
also positive semidefinite.
Its maximum eigenvalue can be bounded via its trace
\begin{align*}
\trace{\MM_{i}}
& = \frac{\delta}{\er_{i_j}} \trace{\left(\YY + \kappa \XX\right)^{-1/2}
\YY_{i_j} \left(\YY + \kappa \XX\right)^{-1/2}}\\
& = \delta \frac{\trace{\left(\YY + \kappa \XX\right)^{-1} \YY_{i_j}}}
{\trace{\XX^{-1} \YY_{i_j}}}\\
& \leq \frac{\delta}{\kappa}.
\end{align*}
Where the last inequality follows from $\left(\YY + \kappa \XX\right)^{-1} 
\preceq \left( \kappa \XX\right)^{-1} = \frac{1}{\kappa} \XX^{-1}$.
It can also be checked that $\expct{i_j}{\frac{\delta}{\er_{i_j}}}  \YY_{i_j}
= \frac{\delta}{s} \YY$, therefore
\[
\expct{i_1 \ldots i_t}{\sum_{j  = 1}^{t} \MM_j}
= \left(\YY + \kappa \XX\right)^{-1/2} \YY \left(\YY + \kappa \XX\right)^{-1/2}.
\]
This gives $\mu_{\max} = 1$, but $\mu_{\min}$ can still be as low as
$\frac{1}{1 + \kappa}$.
Note however that $\ZZ$ is formed by adding $\XX$ to the result.
Therefore, to improve the bounds we introduce $\delta^{-1} \kappa$ more matrices
each equaling to $\delta \left(\YY + \kappa \XX\right)^{-1/2}
\XX \left(\YY + \kappa \XX\right)^{-1/2}$.
As $\left(\YY + \kappa \XX\right)^{-1/2} \preceq \frac{1}{\kappa} \XX^{-1}$,
the maximum eigenvalue in each of these is also at most $\frac{\delta}{\kappa}$.
They on the other hand gives $\expct{}{\sum_{k} \MM_k} = \II$, and therefore
$\mu_{\min} = \mu_{\max} = 1$.

Invoking Lemma~\ref{lem:matrixChernoff} with $R = \delta^{-1} \kappa$
then gives that when $\kappa \defeq c \delta \log{n}$, we have that the eigenvalues
of $\sum_{k} \MM_k$ are between $\frac{1}{2}$ and $2$ with high probability.
Rearranging using the fact that the samples taken equals to
$\ZZ + (\kappa - 1)\XX$ gives
\[
\frac{1}{2} \left( \YY + \kappa \XX \right)
\preceq \ZZ + (\kappa - 1)\XX
\preceq 2 \left( \YY + \kappa \XX \right).
\]

The $\XX$ terms can then be removed using the fact that
$\bvec{0} \preceq \XX \preceq \YY$, giving
\[
\frac{1}{2} \YY
\preceq \frac{1}{2} \left( \YY + \kappa \XX \right)
\preceq \ZZ + (\kappa - 1)\XX
\preceq \kappa \ZZ,
\]
for the lower bound, and 
\[
\ZZ \preceq \ZZ + (\kappa - 1)\XX
\preceq 2 \left( \YY + \kappa \XX \right)
\preceq 2 (\kappa + 1) \YY,
\]
for the upper bound.
Recalling that $\kappa = c \delta \log{n}$ then gives the bound.
\QED

Invoking this with $\delta = \Oh(1)$, and analyzing the amplification i
error caused by sampling too many off-tree edges in the same way as
Lemma~\ref{lem:randPrecon} then gives Lemma~\ref{lem:spectralBound}.

\section{Propagation and Removal of Errors}

As all intermediate solutions in our algorithms contain errors,
we need to check that these errors propagate in a natural way
across the various combinatorial transformations.
We do this by adapting known analyses of the recursive
preconditioning framework~\cite{SpielmanTengSolver}
and Steiner tree preconditioners~\cite{MaggsMOPW05,Koutis:thesis}
to a vector convergence setting.
We also check that it suffices to perform all intermediate
computations to a constant factor relative errors by
showing an outer loop that reduces this error to $\epsilon$
in $O(\log(1/\epsilon))$ iterations.

\subsection{Partial Cholesky Factorization}

\greedyElimination*

\Proof
The greedy elimination procedure from Section 4.1. of~\cite{SpielmanTengSolver} gives a factorization of $\LL_{H}$ into
\[
\LL_{H}
=
\UU^T \left(
\begin{array}{cc}
\II & 0 \\
0 & \LL_{H'}
\end{array}
\right)
\UU,
\]
where $\LL_{H'}$ has $O(m')$ vertices and edges and
for any vector $\yy$, both $\UU^{-T} \yy$ and $\UU^{-1} \yy$
can be evaluated in $O(n)$ time.
It can also be checked that this elimination routine preserves
the stretch of off-tree edges, giving a tree $T'$ as well.

For notational simplicity, we will denote the block-diagonal matrix
with $\II$ and $\LL_{H'}$ as $\PP$.
Note that $\II$ and $\LL_{H'}$ act on orthogonal subspaces
since their support are disjoint and solving a linear system in $\II$ is trivial.
This means that making one call to $\textsc{Solve}$ with $(H', T', \er')$
 plus $O(n)$ overhead gives solver routine for $\PP$.
More specifically, we have access to a routine $\textsc{Solve}_{\PP}$ such that
for any vector $\bb'$, $\xx' = \textsc{Solve}_{\PP}(\bb', \epsilon)$ obeys:
\[
\nbr{\xx' - \PP^{\dag} \bb'}_{\PP}
\leq \epsilon \nbr{\PP^{\dag} \bb'}_{\PP}.
\]

We can then check incorporating $\UU^{-1}$ and $\UU^{-T}$
the natural way preserves errors.
Given a vector $\bb$, we call $\textsc{Solve}_{\PP}$ with
the vector $\bb' = \UU^{-T} \bb$, and return $\xx = \UU^{-1} \xx'$.
Substituting the error guarantees above gives
\[
\nbr{\UU \xx - \PP^{\dag} \UU^{-T} \bb}_{\PP}
\leq \epsilon \nbr{\PP^{\dag} \UU^{-T} \bb}_{\PP}.
\]
Incorporating $\LL_H^{\dag} = \UU^{-1} \PP^{\dag} \UU^{-T}$ then gives
\[
\nbr{\UU \left( \xx - \LL_H^{\dag} \right)\bb}_{\PP}
\leq \epsilon \nbr{\bb}_{\LL_H^{\dag}},
\]
which simplifies to
\[
\nbr{\xx - \LL_H^{\dag} \bb}_{\LL_H}
\leq \epsilon \nbr{\LL_H^{\dag} \bb}_{\LL_H}.
\]
\QED

\subsection{Transfer of Errors}

\errorTransfer*

\Proof
We first check that the RHS terms are equal to each other
by switching the matrix norms.
\[
\norm{\LL_{G}^{\dag} \Pi^T \PPi_1^T \bbhat}_{\LL_G}
= \norm{ \Pi^T \PPi_1^T \bbhat}_{\LL_{G}^{\dag}}
= \norm{\bbhat}_{ \PPi_1 \Pi \LL_{G}^{\dag}  \Pi^T \PPi_1^T }
= \norm{\PPi_1 \PPi \LL_{G}^{\dag}  \PPi^T \PPi_1^T\bbhat}
_{ \left(\PPi_1 \PPi \LL_{G}^{\dag}  \PPi^T \PPi_1^T\right)^{\dag}}.
\]

A similar manipulation of the LHS gives:
\[
\norm{\xxhat - \PPi_1 \PPi \LL_{G}^{\dag}  \PPi^T \PPi_1^T \bbhat}
_{\left(\PPi_1 \PPi \LL_{G}^{\dag}  \PPi^T \PPi_1^T\right)^{\dag}}
= \norm{\PPi_1 \PPi  \left( \xx - \LL_{G}^{\dag}  \PPi^T \PPi_1^T \bbhat \right)}
_{\left(\PPi_1 \PPi \LL_{G}^{\dag}  \PPi^T \PPi_1^T\right)^{\dag}}.
\]
Note that
$\left(\PPi_1 \PPi \LL_{G}^{\dag}  \PPi^T \PPi_1^T\right)^{\dag}$
is the Schur complement of $\LL_G$ on its rank space
onto the column space of $\PPi_1 \PPi$.
As the Schur complement quadratic form gives the minimum energy
over all extensions of the vector w.r.t. the original quadratic form,
we have:
\[
\norm{\PPi_1 \PPi  \left( \xx - \LL_{G}^{\dag}  \PPi^T \PPi_1^T \bbhat \right)}
_{\left(\PPi_1 \PPi \LL_{G}^{\dag}  \PPi^T \PPi_1^T\right)^{\dag}}
\leq 
\norm{\xx - \LL_{G}^{\dag} \PPi^T \PPi_1^T \bbhat}_{\LL_{G}}.
\]
which when combined with the equality for the RHS completes
the result.

\subsection{Preconditioned Richardson Iteration}

\begin{restatable}{lemma}{preconRichardson}
\label{lem:preconRichardson}
If $\AA$, $\BB$ are matrices such that $\AA \preceq \BB \preceq 2 \AA$
and $\textsc{Solve}_{\BB}$ is a routine such that for any vector $\vecb$
we have $\nbr{\textsc{Solve}_{\BB}(\vecb)- \mata^{\dag} \vecb}_{\BB}
\leq \frac{1}{5} \nbr{\mata^{\dag} \vecb}_{\BB}$.
There is a routine $\textsc{Solve}_{\AA,\epsilon}$ which runs in $\Oh(c_{\alpha}\log(\frac{1}{\epsilon}))$ iterations with the guarantee that for any vector $\vecb$ we have $\nbr{\textsc{Solve}_{\AA,\epsilon}(\vecb)- \mata^{\dag}\vecb}_{\AA} \leq \epsilon \nbr{\AA^{\dag} \vecb}_{\AA}$.
Each iteration involves one call to $\textsc{Solve}_{\BB}$,
a matrix-vector multiplication involving $\AA$ and operations on vectors.
\end{restatable}

\begin{figure}
\vskip 0.2in
\noindent
\fbox{
\begin{minipage}{6in}

\begin{enumerate}

\item $\xx = \textsc{Solve}_{\BB} \left(  \bb  \right)$
\item Let $t= \log_{\alpha}(\frac{1}{\epsilon})$. For $i=0...t$ \\
$\yy = \textsc{Solve}_\mata \left(  \vecb - \mata \vecx  \right)$\\
$\xx = \vecx +  \vecy$
\item Return $\vecx$
\end{enumerate}

\end{minipage}
}
\vskip 0.2in
\caption{Preconditioned Richardson Iteration}
\label{fig:preconRichardson}
\end{figure}

\Proof
A pseudocode of the routine $\textsc{Solve}_{\matb}$ is given in Figure~\ref{fig:preconRichardson}.
It suffuces to show that each iteration,
$\nbr{\vecx - \mata^{\dag} \vecb}_{\mata}$ decreases by a constant factor.

We will use $\xx'$ to denote the solution vector produced for the
next iteration.
As our convergence is in terms of distance to the exact solution,
it is convenient to denote the current error
using $\rr = \vecx - \mata^{\dag} \vecb$.

Applying the triangle inequality to the new error gives:
\[
\nbr{\xx' - \AA^{\dag} \vecb}_{\AA}
= \nbr{\xx +  \yy- \AA^{\dag} \vecb}_{\AA}
\leq \nbr{\xx - \AA^{\dag}\bb + \BB^{\dag} \left(\bb - \AA \xx \right)}_{\AA}
+ \nbr{\yy - \bb^{\dag} \left( \bb - \AA \xx  \right) \rr}_{\AA}.
\]
If $\bb$ is in the column space of $\AA$ and $\BB$,
$\bb - \AA \xx = \AA ( \AA^{\dag} \bb - \xx) = -\AA \rr$.
As the error is measured in the $\AA$-norm,
we can make this substitution, giving:
\[
\nbr{\xx' - \AA^{\dag} \vecb}_{\AA}
\leq  \nbr{\left( \II - \BB^{\dag} \AA \right) \rr}_{\AA}
+ \nbr{\yy - \BB^{\dag} \AA \rr}_{\AA}.
\]
The first term equals to
\[
\sqrt{\rr^T \AA^{1/2} \left( \II - \AA^{1/2} \BB^{\dag} \AA^{1/2} \right)^2 \AA^{1/2} \rr}
\]
Rearranging $\AA \preceq \BB \preceq 2 \AA$ gives
$0 \preceq \II - \AA^{1/2} \BB^{\dag} \AA^{1/2} \preceq \frac{1}{2} \II$,
which means the first term can be bounded by $\frac{1}{2} \nbr{\rr}_{\AA}$.

The second term can be bounded using the guarantees of
$\textsc{Solve}_{\AA}$ and the bounds between $\AA$ and $\BB$:
\[
\nbr{\yy - \BB^{\dag} \AA \rr}_{\AA}
\leq \nbr{\yy - \BB^{\dag} \AA \rr}_{\BB}
\leq \alpha \nbr{\rr}_{\BB}
\leq 2 \alpha \nbr{\rr}_{\AA}.
\]
Summing these two terms gives $\nbr{\xx' - \AA^{\dag} \vecb}_{\AA}
\leq \frac{9}{10} \nbr{\xx - \AA^{\dag} \vecb}_{\AA}$, 
and therefore the convergence rate.
\QED

\end{appendix}

\end{document}